\documentclass[12pt]{iopart}

\usepackage{array}
\usepackage{amsmath}
\usepackage{amsgen}
\usepackage{amsfonts}
\usepackage{amsbsy}
\usepackage{amssymb}

\usepackage{braket}

\usepackage{graphicx}
\usepackage{dcolumn}
\usepackage{bm,textcomp}

\newcommand{\posi}[1]{\ket{\alpha_{#1}}} 
\newcommand{\posr}[1]{\ket{\widetilde{\alpha}_{#1}}} 
\newcommand{\poss}[1]{\ket{\widetilde{\alpha}^S_{#1}}} 
\newcommand{\pose}[1]{\ket{\widetilde{\alpha}^E_{#1}}} 

\usepackage[figuresleft]{rotating}
\usepackage{units}

\def\idty{{\leavevmode\rm 1\mkern -5.4mu I}} 

\begin{document}

\title[Experimental Simulation and Limitations of QWs with Trapped Ions]{Experimental Simulation and Limitations of Quantum Walks with Trapped Ions}

\author{R Matjeschk$^{1,2}$, Ch Schneider$^{2,3}$, M Enderlein$^{2,3}$, T Huber$^{2,3}$, H Schmitz$^{2}$, J Glueckert$^{2}$ and T Schaetz$^{2,3}$}
\address{$^{1}$ Institut f\"ur Theoretische Physik, Leibniz Universit\"at Hannover, Appelstr. 2, 30167 Hannover, Germany}
\address{$^{2}$ Max-Planck-Institut f\"ur Quantenoptik, Hans-Kopfermann-Str. 1, 85748 Garching, Germany}
\address{$^{3}$ Albert-Ludwigs-Universit\"at Freiburg, Physikalisches Institut, Hermann-Herder-Str. 3, 79104 Freiburg, Germany}
\ead{robert.matjeschk@itp.uni-hannover.de}

\date{\today}

\begin{abstract}
We examine the prospects of discrete quantum walks (QWs) with trapped ions. In particular, we analyze in detail the limitations of the protocol of Travaglione and Milburn (PRA 2002) that has been implemented by several experimental groups in recent years. Based on the first realization in our group (PRL 2009), we investigate the consequences of leaving the scope of the approximations originally made, such as the Lamb--Dicke approximation. We explain the consequential deviations from the idealized QW for different experimental realizations and an increasing number of steps by taking into account higher-order terms of the quantum evolution. It turns out that these become dominant after a few steps already, which is confirmed by experimental results and is currently limiting the scalability of this approach. Finally, we propose a new scheme using short laser pulses, derived from a protocol from the field of quantum computation. We show that the new scheme is not subject to the above-mentioned restrictions, and analytically and numerically evaluate its limitations, based on a realistic implementation with our specific setup. Implementing the protocol with state-of-the-art techniques should allow for substantially increasing the number of steps to 100 and beyond and should be extendable to higher-dimensional QWs.
\end{abstract}


\maketitle

\section{Introduction}
Random walks are powerful models that allow to describe, understand
and make use of stochastic processes occuring in a wide variety of
areas \cite{Barber1970, Berg1993}. Models describing related
processes in the quantum world are called quantum walks (QWs)
\cite{Aharonov1993}.

An implementation of a discrete (-time and -space) random walk on a
line requires two basic operations. The coin operation, with the
random outcome of Heads or Tails, is followed by the shift operation
to the left or right, depending on the outcome of the coin toss.
After $N$ steps the walker will therefore have followed randomly one
out of many possible paths, with the probability for its location
being given by a binomial distribution centered around the starting
point. The average displacement of the walker, i.e. the standard
deviation of that distribution, increases with the square root of
$N$. The quantum mechanical version replaces the probabilistic coin
toss by a deterministic operation. It prepares the quantum coin in
an (equal) superposition of Heads and Tails. As a consequence, the
walker performs the conditional step in both directions
simultaneously. The walker follows all paths during this
deterministic (and thus reversible) process, allowing for
constructive and destructive interferences at subsequent crossings.
The probability distribution of the position of the walker is due to
these interferences substantially different from a binomial one. In
particular, the average displacement of the walker scales linearly
in $N$.

Quantum walks have been thoroughly investigated theoretically and
several applications for QWs have been proposed, for example in
terms of quantum computing \cite{Kempe2003}. Many classical
algorithms in computer science make use of random walks for sampling
purposes. Algorithms of that kind might get substantially speeded up
by quantum versions of the random walk, where all possible paths are
tested in parallel, potentially providing a similar gain as the
prominent example of Grover's search algorithm \cite{Grover2001}. In
addition, QWs can be interpreted as the one-particle sector of a
quantum cellular automaton, which is a fundamental model of a
quantum computer \cite{Gross2009}. Furthermore it has been shown
that QWs themselves are suitable for universal quantum computation
\cite{Childs2009} and different aspects of quantum information
processing \cite{Ambainis2007, Shenvi2003, Shikano2010}.

In a different context, QWs can be exploited as prototype models for
intriguing transport processes in nature. One examples is the energy
transfer in photosynthesis with an efficiency of close to 100\%
\cite{Engel2007, Mohseni2008}, a performance that is not achievable
classically. Other examples are the creation of molecules in
interacting QWs \cite{Ahlbrecht2011a} and effects like Anderson
localization and diffusive scaling in disordered QWs
\cite{Ahlbrecht2011b, Ahlbrecht2011}. Here, QWs might be suited for
experimental quantum simulations to provide deeper insight into
complex quantum dynamics. Additionally, even relativistic effects
can be considered \cite{Chandrashekar2010}.

Promising attempts at their implementation have been performed for
the discrete and the continuous versions of QWs. Important aspects
of QWs have been realized in a nuclear magnetic resonance experiment
\cite{Ryan2005} using the internal degrees of freedom of molecules
to span the coin and position space. An implementation based on
neutral atoms in an optical lattice \cite{Mandel2003, Duer2002} has
resulted in an experiment \cite{Karski2009} where the lattice sites
in a standing wave of light span the position space of the
walker/atom, two electronic states encode the two coin states and a
state-dependent optical force provides the conditional shift. Other
proposals consider an array of microtraps illuminated by a set of
microlenses \cite{Eckert2005}, Bose--Einstein condensates
\cite{Chandrashekar2006} and atoms in cavities
\cite{Bouwmeester1999}. Photons have mimicked single walkers on the
longitudinal modes of a linear optical resonator
\cite{Bouwmeester1999} and in a loop of a split optical fibre
\cite{Schreiber2010}. Single \cite{Perets2008} and two
time-correlated photons \cite{Peruzzo2010, Schaetz2011} have
recently been travelling and interfering in a lattice of optical
waveguides. Travaglione and Milburn \cite{Travaglione2002} proposed
a scheme for trapped ions to transfer the high operational
fidelities \cite{Wineland2011} obtained in quantum information
processing into the field of QWs. While coin states and steps are
operated similar to the atoms in the optical lattice, the position
is encoded in the motional degree of freedom of the ion(s), which
oscillate in a quantized harmonic trapping potential. The
proof-of-principle has been performed by our group by the
implementation of a discrete, asymmetric QW of one trapped ion along
a line in phase space \cite{Schmitz2009a}. Recently, the proposal
has been theoretically refined \cite{Xue2009} and experimentally
extended to an increased number of steps \cite{Zaehringer2010}.

All of the above-described systems and their related protocols of
implementation are severely limited in the total number of steps due
to a lack of operation fidelities or even fundamental restrictions.
However, a larger number of precisely performed steps is the crucial
prerequisite to exploit QWs for the envisioned applications. For the
case of trapped ions, the limit of coherent displacements to states
inside the Lamb--Dicke regime has already been foreseen
\cite{Travaglione2002}, experimentally observed in a different
context \cite{McDonnell2007, Poschinger2010}, and confirmed by us
\cite{Schmitz2009} and others \cite{Zaehringer2010}.

In this paper we substantially extend the description of the
experimental implementation of the asymmetric QW with three steps
\cite{Schmitz2009}. We carefully analyze the effects that arise when
approaching the fundamental limitations of the proposed protocol
\cite{Travaglione2002} after the related, severely restricted number
of steps for different step sizes. We consider higher-order terms to
the soon overstrained approximation building on the work of
\cite{McDonnell2007}. In parallel, we experimentally confirm the
essentials by further investigating our results \cite{Schmitz2009},
which already lead into a regime where the refined theory is
required. Finally, we develop a novel protocol for a QW, based on a
scheme from the field of quantum information processing using
photon kicks \cite{Garc'ia-ripoll2003, Garc'ia-ripoll2005}, to
overcome these restrictions and to allow in principle for hundred(s)
of steps, extendable to QWs in higher dimensions.

The paper is structured as follows: In section \ref{theory} we give 
a theoretical description of the QW as it has been realized in our
experiment, similar to the original proposal \cite{Travaglione2002},
and analyze issues concerning non-orthogonality of the position
states. In section \ref{implementation} we describe the experimental
method of realizing the necessary operations of the QW and analyze
the limitation of the position space to the LDR of the optical
dipole force. In section \ref{detection} we describe the
experimental methods of the ion state detection. In section
\ref{realization} we describe the experimental procedure, in
particular the determination of the relevant parameters. In section
\ref{conclusion} we summarize the results and the limitations of the
implementation of the QW. Finally, in section \ref{outlook}, we
propose the implementation of the shift operator with short laser
pulses (photon kicks) and the extension to higher dimensions.

\section{Theoretical considerations}\label{theory}
In the following we give a theoretical description of the discrete
QW on a line as it is realized in our proof-of-principle experiment
for the first three steps.

\begin{figure}
\centering
\includegraphics[width=12cm]{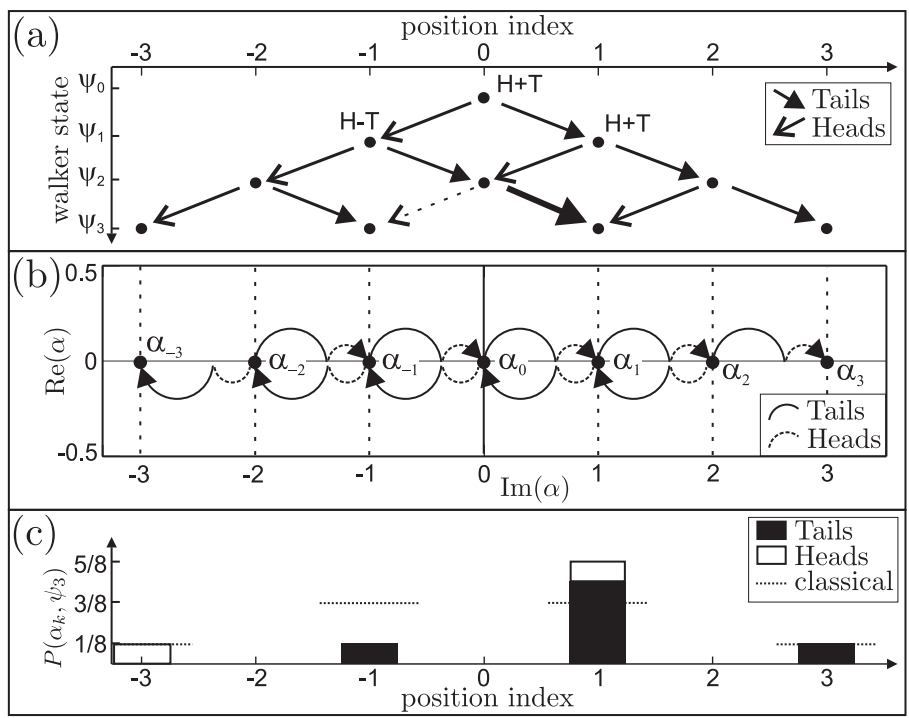}
\caption{The asymmetric QW on a line. (a) Schematic of the QW. Starting at the position with index $0$, the walker tosses a coin and does a shift to the position index $1$ ($-1$) for the coin showing Tails ($T$) (Heads ($H$)). If the coin is in a superposition of $T$ and $H$, the walker shifts into both directions simultaneously, taking the part related to $T$ to position $1$ and the one related to $H$ to position $-1$. Further coin tosses from $T$ ($H$) cause superposition states with different relative phases, that is, $H+T$ and $H-T$ respectively. The relative phases allow for interferences of the wave function between different paths. The first interference occurs during the third step, taking the walker from the state $\psi_2$ to $\psi_3$.
At position $0$ the coin toss results in constructive (destructive) interference for $T$($H$), illustrated by the bold (dashed) arrow for the subsequent shift. (b) We implement the QW with a trapped ion, where the position space is encoded into the co-rotating phase space ($\text{Re}(\alpha)$, $\text{Im}(\alpha)$) \eqref{pr_def} of the axial normal mode of motion. The positions $k$ are represented by coherent states $\ket{\alpha_k}$, which are aligned along a line in the co-rotating phase space. Two electronic (hyperfine) states of the ion encode the coin states.
The transition from position $\ket{\alpha_k}$ to $\ket{\alpha_{k\pm1}}$ is achieved via two subsequent displacements, each followed by a pulse exchanging the the coin states (See figure \ref{kombipuls}).
(c) Probability distribution of the walker in position space after three steps, under the assumption that the position states are orthogonal. The black (white) filled boxes represent the contributions of the wave function related to $\ket{T}$ ($\ket{H}$). The asymmetry between the position state probabilities $P(\alpha_1,\psi_3)=\bra{\psi_3}\bigl(\idty\otimes\ket{\alpha_1}\bra{\alpha_1}\bigr)\ket{\psi_3}$ and $P(\alpha_{-1},\psi_3)$ is due to the interferences indicated via the bold and dashed arrows in (a).
}
 \label{schrittidee}
\end{figure}

Consider a Hilbert space vector
\begin{equation}\ket{\psi} \in \mathbb{H} = \mathbb{H}_{coin} \otimes \mathbb{H}_{motion}.
\end{equation}
$\mathbb{H}_{coin}$ denotes the coin space with basis states
\begin{equation}
\ket{H} = \begin{pmatrix}1 \\ 0 \end{pmatrix}, \quad \ket{T} = \begin{pmatrix}0 \\ 1 \end{pmatrix},
\end{equation}
encoding the coin states, Heads and Tails. $\mathbb{H}_{motion}$ is
the infinite dimensional phase space, related to a harmonic
oscillator. We encode the discrete positions as coherent states
\begin{equation}\label{alphadef}
\ket{\alpha_k} = e^{-\lvert \alpha_k \rvert^2/2} \sum^{\infty}_{n=0} \frac{\alpha^n_k}{\sqrt{n!}} \ket{n},
\end{equation}
where $k \in \mathbb{Z}$ and $\alpha_k = k \cdot \Delta \alpha$ with
$\Delta \alpha \in \mathbb{C}$. The states $\ket{n}$ denote the
(orthonormal) Fock states. For the QW the distance
$\lvert\Delta\alpha\rvert$ between neighbouring positions in phase
space is of importance, whereas the argument of the complex number
$\Delta\alpha$ can be chosen to be constant for all steps and
therefore is irrelevant.

Concerning the notation of the position states, we will use the
following convention. Ideally, the position states are coherent
motional states $\posi{k}$, as described above. In the experiment
the position states will contain a small amount of motional
squeezing (Sect. \ref{stepsizecalibration}). These states will be
denoted as $\posr{k}$. Additionally, whenever necessary, we will
distinguish between position states generated in a numerical
simulation, $\poss{k}$, and experimentally, $\pose{k}$. Further, we
will generally use the superindices $^S$ and $^E$ to distinguish
between simulation and experiment, whenever necessary. For the
simulation we will use 3SB (See sect. \ref{dynamicsdescription}), if
not stated differently.

Ideally, the initial state of the QW is chosen to be
\begin{equation}
\ket{\psi_0} = \ket{T}\otimes\ket{\alpha_0=0}.
\end{equation}
Each step of the QW is described by the subsequent application of
the coin operator $C$ and the shift operator $S$. Thus the state
after $N$ steps is given by $\ket{\psi_N}=(S\cdotp
C)^N\ket{\psi_0}$.

The coin operator $C$ is defined as
\begin{equation}\label{coinop}
C=R\left(\frac{\pi}{2},\phi\right)=\frac{1}{\sqrt{2}}\begin{pmatrix} 1 & e^{i\phi}\\ -e^{-i\phi} & 1 \end{pmatrix} \otimes \idty_{motion},
\end{equation}
according to
\begin{equation}\label{rfmatrix}
R \left( \theta, \phi \right) = \begin{pmatrix} \cos\left(\theta/2\right) & e^{i\phi}\sin\left(\theta/2\right) \\ -e^{-i\phi}\sin\left(\theta/2\right) & \cos\left(\theta/2\right) \end{pmatrix} \otimes \idty_{motion}.
\end{equation}

From the initial state $\ket{\psi_0}$, the operator $C$ with $\phi$
being arbitrary, but equal for every application of $C$, leads to an
asymmetric QW (Figure \ref{schrittidee}). A symmetric QW can be
realized with the coin operator for the first step being $R
\left(\pi/2, \phi \right)$ and for all following steps being
$R\left(\pi/2, \phi+\pi/2 \right)$ (with $\phi$ arbitrary). In that
case, the first coin toss can be interpreted as the initialization
of the coin state such that all following coin tosses act
symmetrically on it.

The shift operator $S$ is defined as
\begin{equation}\label{stepop}
S =\ket{T}\bra{T} \otimes D(\Delta\alpha) + \ket{H}\bra{H} \otimes D(-\Delta\alpha),
\end{equation}
with $D(\Delta\alpha)=\text{exp}(\Delta\alpha \cdotp a^\dagger - \Delta\alpha^* \cdotp a)$ being the displacement operator and $a^\dagger$,$a$ the corresponding raising and lowering operators.

\begin{figure}
\centering
\includegraphics[width=16cm]{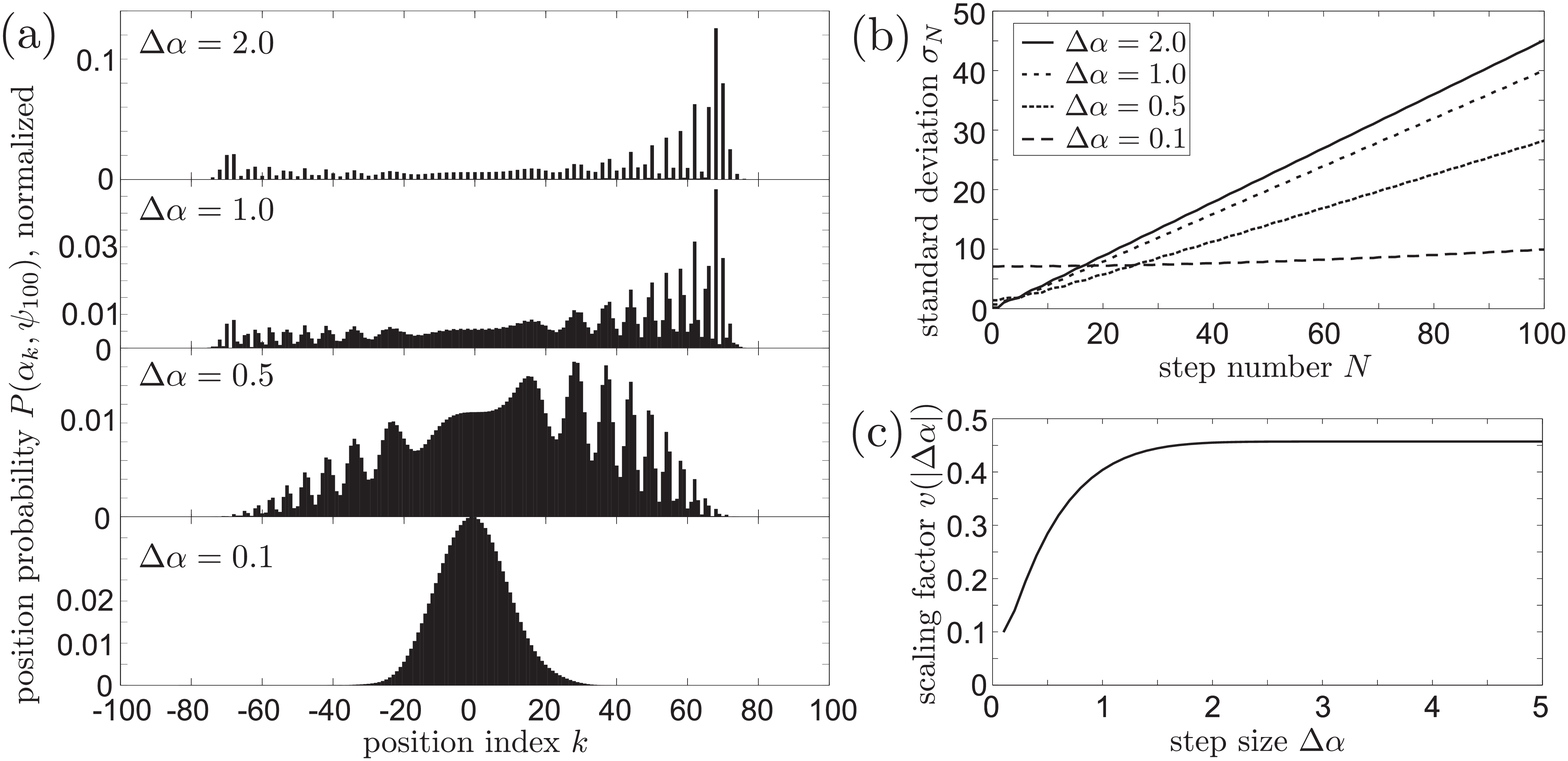}
\caption{Characteristics of the probability distribution for asymmetric QWs of different step sizes $\lvert\Delta\alpha\rvert$. (a) Position probabilities $P\left(\alpha_k,\psi_{100}\right)$ after $N=100$ steps. The probabilities are normalized to $\sum_k P\left(\alpha_k,\psi_{100}\right)$=1. For $\lvert\Delta\alpha\rvert \geq 2$ the position states are approximately orthogonal and therefore the probability distribution shows the shape of an orthogonal QW \cite{Kempe2003}. In particular, the probabilities for positions $\ket{\alpha_k}$ with odd index $k$ vanish. For $\lvert\Delta\alpha\rvert = 1$ the probability distribution is smeared out and all position states are populated.
However, the characteristic peaks around $k=70$ are still prominent. For $\lvert\Delta\alpha\rvert = 0.5$ the probability distribution does not feature the main peak around $k=70$ any more. For $\lvert\Delta\alpha\rvert = 0.1$ the probability distribution approaches a Gaussian shape. (b) Standard deviation $\sigma_N$ \eqref{standarddeviation} in dependence on $N$ for QWs with different step sizes $\lvert\Delta\alpha\rvert$. After a few initial steps the standard deviation scales linearly in $N$, i.e. $\sigma_N=v(\lvert\Delta\alpha\rvert)\cdotp N$. For $\lvert\Delta\alpha\rvert=0.1$ already the initial state $\ket{\psi_0}$ of the walker is considerably spread out over the position states $\ket{\alpha_k}$ such that the standard deviation differs significantly from zero. The linear scaling becomes evident for $N\gtrsim60$. (c) Scaling factor $v(\lvert\Delta\alpha\rvert)$ of the standard deviation for different step sizes $\lvert\Delta\alpha\rvert$. For $\lvert\Delta\alpha\rvert\geq2$ the scaling factor is larger than $99\%$ of the asymptotic value ($v(\lvert\Delta\alpha\rvert\rightarrow\infty)=0.457$), which is the scaling of a QW with orthogonal position states. In our experiment we set the step size to $\lvert\Delta\alpha\rvert\approx1$, which results in a scaling factor of $89\%$ of the asymptotic value.
}
 \label{qw_schmelz}
\end{figure}

In contrast to a typical QW\footnote{In the sense that the position
states are orthogonal, which has to our knowledge been assumed in
the vast majority of publications concerning QWs, so far.} the
position states $\ket{\alpha_k}$ are not orthogonal. The step size
$\lvert\Delta\alpha\rvert$ determines the overlap of the position
states, $\braket{\alpha_k|\alpha_l} =
\text{exp}(-(k-l)^2\lvert\Delta\alpha\rvert^2 / 2)$. If the state of
the walker after $N$ steps is
$\ket{\psi_N}=\sum^N_{k=-N}\left(c^T_k\ket{T}\ket{\alpha_k}+c^H_k\ket{H}\ket{\alpha_k}\right)$,
the probability of finding the walker in position $\ket{\alpha_L}$
is given by

\begin{equation}
\begin{split}
P(\alpha_L,\psi_N)=&\bra{\psi_N}\bigl(\idty\otimes\ket{\alpha_L}\bra{\alpha_L}\bigr)\ket{\psi_N}\\
=&\left|\sum^N_{k=-N}c^H_k\cdotp\braket{\alpha_k|\alpha_L}\right|^2+\left|\sum^N_{k=-N}c^T_k\cdotp\braket{\alpha_k|\alpha_L}\right|^2.
\end{split}
\end{equation}
Thus only if the step size $\lvert\Delta\alpha\rvert$ is large
enough such that the overlap between different position states
remains negligible, the above probability is given by the
coefficients $c^H_L$ and $c^T_L$ only. In figure \ref{qw_schmelz}
the probability distributions after 100 steps for QWs with different
step sizes $\lvert\Delta\alpha\rvert$ are illustrated. We find that
for $\lvert\Delta\alpha\rvert\geq2$, where the position states
contain a negligible overlap of
$\lvert\braket{\alpha_k|\alpha_{k+1}}\rvert^2\leq e^{-4}$, the
probability distribution shows the shape of an orthogonal QW
\cite{Kempe2003}. For smaller values of $\lvert\Delta\alpha\rvert$
the probability distributions are smeared out due to the increased
overlaps between the position states. As $\lvert\Delta\alpha\rvert$
approaches zero, the probability distribution approaches a Gaussian
shape. The mean distance of the walker from the origin, which is
given by the standard deviation
\begin{equation}\label{standarddeviation}
\sigma_N=\sqrt{\braket{k^2}_N-\braket{k}^2_N}
\end{equation}
with
\begin{equation}
\braket{f(k)}_N=\frac{\sum^N_{k=-N}f(k)\cdotp P(\alpha_k,\psi_N)}{\sum^N_{k=-N}P(\alpha_k,\psi_N)}
\end{equation}
for any function $f(k)$, grows slower with $N$ for smaller values of
$\lvert\Delta\alpha\rvert$. This is because the lower the value of
$\lvert\Delta\alpha\rvert$, the less the shift operator $S$ actually
changes the state of the walker. However, for every realization of
$\lvert\Delta\alpha\rvert$ the average distance of the walker from
the origin scales linearly with the step number $N$, i.e. $\sigma_N
\equiv v(\Delta\alpha)\cdotp N$, after a certain number of steps, as
depicted in figure \ref{qw_schmelz}b. The asymptotic scaling for the
limit $\lvert\Delta\alpha\rvert\rightarrow0$ has not been
investigated yet.

In our experiment we set the step size to
$\lvert\Delta\alpha\rvert\approx1$. In this case the overlaps
between the position states amount to
$\lvert\braket{\alpha_k|\alpha_{k+1}}\rvert^2=1/e$. The
characteristic of the probability distribution after 100 steps
(Figure \ref{qw_schmelz}a) is still close to that of
$\lvert\Delta\alpha\rvert\geq2$ and thus an orthogonal QW.

Ideally, with the initial state $\ket{\psi_0}$ and $\phi=0$ \eqref{coinop} three steps of the asymmetric QW lead to the state
\begin{equation}\label{psi3}
\ket{\psi_3} = \frac{1}{\sqrt{8}} \ket{T}\otimes \Bigl(-2\ket{\alpha_1}+\ket{\alpha_3}-\ket{\alpha_{-1}} \Bigr)
+\frac{1}{\sqrt8} \ket{H}\otimes \Bigl( \ket{\alpha_{-3}} + \ket{\alpha_1} \Bigr).
\end{equation}

The probabilities of finding the walker in the coin state $\ket{H}$
or $\ket{T}$ after three steps are given by $P_H\left(\psi_3\right)
= \bra{\psi_3}\bigl(\ket{H}\bra{H}\otimes \idty\bigr)\ket{\psi_3}$
and $P_T\left(\psi_3\right) =
\bra{\psi_3}\bigl(\ket{T}\bra{T}\otimes \idty\bigr)\ket{\psi_3}$.
Their ratio
\begin{equation}\label{QWratio}
\frac{P_H\left(\ket{\psi_3}\right)}{P_T\left(\ket{\psi_3}\right)} = \frac{1+e^{-8\lvert\Delta\alpha\rvert^2}}{3-e^{-8\lvert\Delta\alpha\rvert^2}}
\end{equation}
amounts to $\approx 1/3$ for $\lvert\Delta\alpha\rvert \geq 1$.

\section{Implementation of the QW}\label{implementation}
\subsection{System and definitions}
For the experimental implementation we confine a single
$^{25}$Mg$^+$ ion in a linear Paul trap \cite{Schaetz2007}. The
motional frequency related to the confinement in the axial direction
of the trap is set to $\omega_z = 2\pi \cdotp 2.13 \text{ MHz}$ and
in the radial directions to $\omega_x \approx \omega_y \approx 2 \pi
\cdot 5 \text{ MHz}$. We define two out of 12 electronic states of
the hyperfine ground state manifolds \cite{Schmitz2009} (Figure
\ref{termschema})
\begin{equation}
\begin{split}
\ket{H}&\equiv\ket{^2S_{1/2}, F=2, m_F=2},\\
\ket{T}&\equiv\ket{^2S_{1/2}, F=3, m_F=3}
\end{split}
\end{equation}
as the coin states. Further, we will use the state
\begin{equation}
\ket{A}\equiv\ket{^2S_{1/2}, F=2, m_F=-2}
\end{equation}
in the detection procedure. To lift the degeneracy within each
hyperfine manifold, we apply a magnetic field inducing a Zeeman
shift with an energy separation related to $\omega_{Zm} \approx
2\pi\cdotp3 \text{ MHz}$ between neighbouring states. The energy
difference (including the hyperfine splitting) between $\ket{H}$ and
$\ket{T}$ amounts to a frequency of $\omega_{coin} = 2\pi \cdotp
1.77 \text{ GHz}$.

Our realization of the QW consists of the application of a sequence
of laser and radio-frequency (RF) pulses to (1) initialize the ion's
electronic and motional state, (2) to implement the QW, and (3) to
readout the final state via photon scattering. The experiments are
repeated on the order of 1000 times for each set of parameters to
obtain the required statistical relevance. A concise discussion of
these tools in a generic context can be found in \cite{Wineland1998}
and \cite{Leibfried2003}.

\subsection{State initialization}
At the beginning of each experiment the ion is prepared in the coin
state $\ket{T}$ with a fidelity $\geq 0.99$ by optical pumping
\cite{Monroe1995}, while the axial mode of motion is cooled close to
the ground state by Doppler cooling ($\overline{n} \approx 10$)
\cite{Wineland1979} and subsequent sideband cooling ($\overline{n} <
0.03$) \cite{Monroe1995}. The phase space of the axial mode of
motion is used to encode the position of the walker. The radial
modes are Doppler cooled ($\overline{n} \approx 4$), which enables a
sufficient decoupling from the axial degree of freedom.

\subsection{The coin operator}\label{coinimplementation}
We drive coherent transitions between the coin states by applying a
radio-frequency field (RF) for a duration $t$ with frequency
$\omega_{coin}$ (Figures \ref{dipolkraftschema} and
\ref{termschema}) \cite{Wineland1998}. This implements the operator
$R \left( \theta, \phi \right)$ \eqref{rfmatrix} with $\theta =
\Omega\cdotp t$ and $\Omega = 2\pi\cdotp100 \text{ kHz}$, the Rabi
frequency of the transition. The phase $\phi$ for the first pulse of
each experimental cycle is arbitrary. For every following pulse the
phase $\phi$ is set identical with respect to the first pulse. The
duration of one $R \left( \pi, \phi \right)$-pulse amounts to
$T_{\pi} = 5 \text{ }\mu\text{s}$. The coherence time (drop of
oscillation contrast below 50\%) for the RF field exceeds one second
\cite{Schmitz2009}. The duration of a single experimental cycle
(without initialization and detection) of the QW with three steps
amounts to 150 $\mu$s (Figure \ref{pulsschema}). Therefore the
dephasing of the coin remains small. Spin-echo sequences
\cite{Wineland1998}, which are included in the QW pulse sequence
(Figure \ref{pulsschema}), further reduce the dephasing.

\subsection{The shift operator}
\begin{figure}
\centering
\includegraphics[width=15cm]{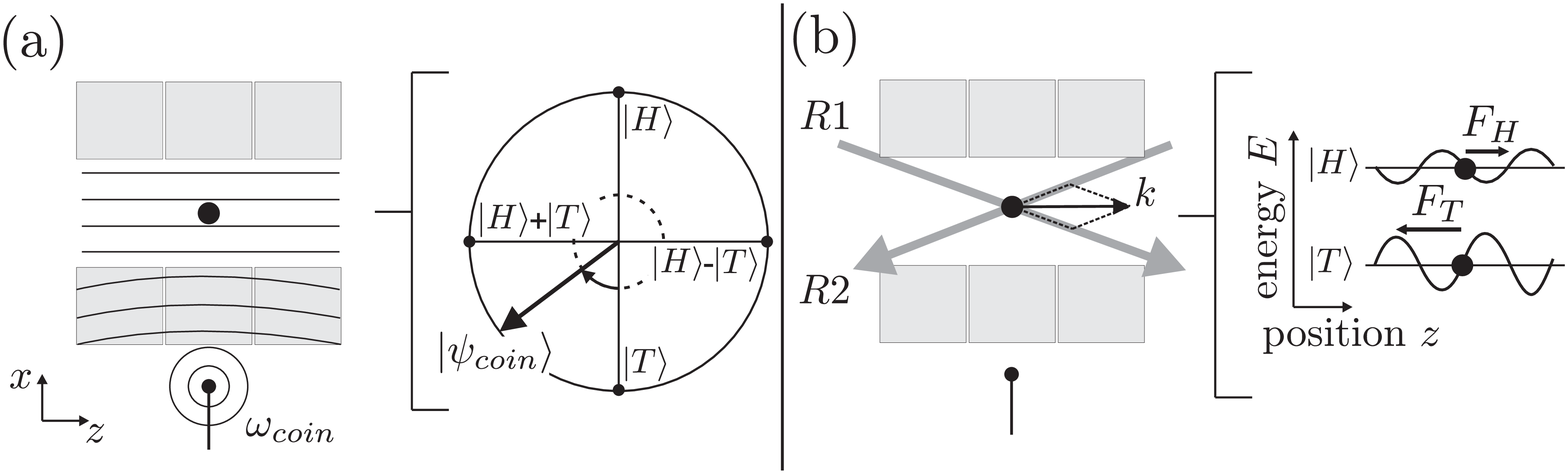}
\caption{Schematic of the operations required for the implementation of the QW. (a) Implementation of the coin operator $R(\theta,\phi)$ \eqref{rfmatrix}. Left: The grey boxes represent a side view of two out of four electrodes of the Paul trap. The black dot between them depicts the trapped ion.  We apply a radio frequency field (RF) via an antenna (below the grey boxes), driving coherent transitions between the coin states $\ket{T}$ and $\ket{H}$. The straight solid lines represent the phase fronts of the RF-field. Right: In our protocol of the QW we apply $R(\theta=\frac{\pi}{2},\phi)$-pulses ($\phi=\text{const.}$), rotating the state vector (Bloch representation) of $\ket{\psi_{coin}}$ by $90^\circ$ from $\ket{T}$ ($\ket{H}$) to $\ket{H}+\ket{T}$ ($\ket{H}-\ket{T}$).
(b) State dependent optical dipole force for the implementation of the shift operator $S$ \eqref{stepop}. Left:
We apply two laser beams, $R1$ and $R2$, with a frequency difference $\omega_1 - \omega_2 =\omega_L = \omega_z-\delta$, perpendicular in polarization and beam direction. The effective wave vector is $\bi{k}=k\bi{e_z}$, pointing into the axial direction $z$. Right: This creates a walking standing wave and related state dependent ac-Stark shifts on the coin states $\ket{H}$ and $\ket{T}$, providing state dependent oscillating forces $F_T$ ($F_H$) (solid sinusoidal lines) acting on the ion in the $z$-direction with frequency $\omega_L$.}
\label{dipolkraftschema}
\end{figure}

\begin{figure}
\centering
\includegraphics[width=10cm]{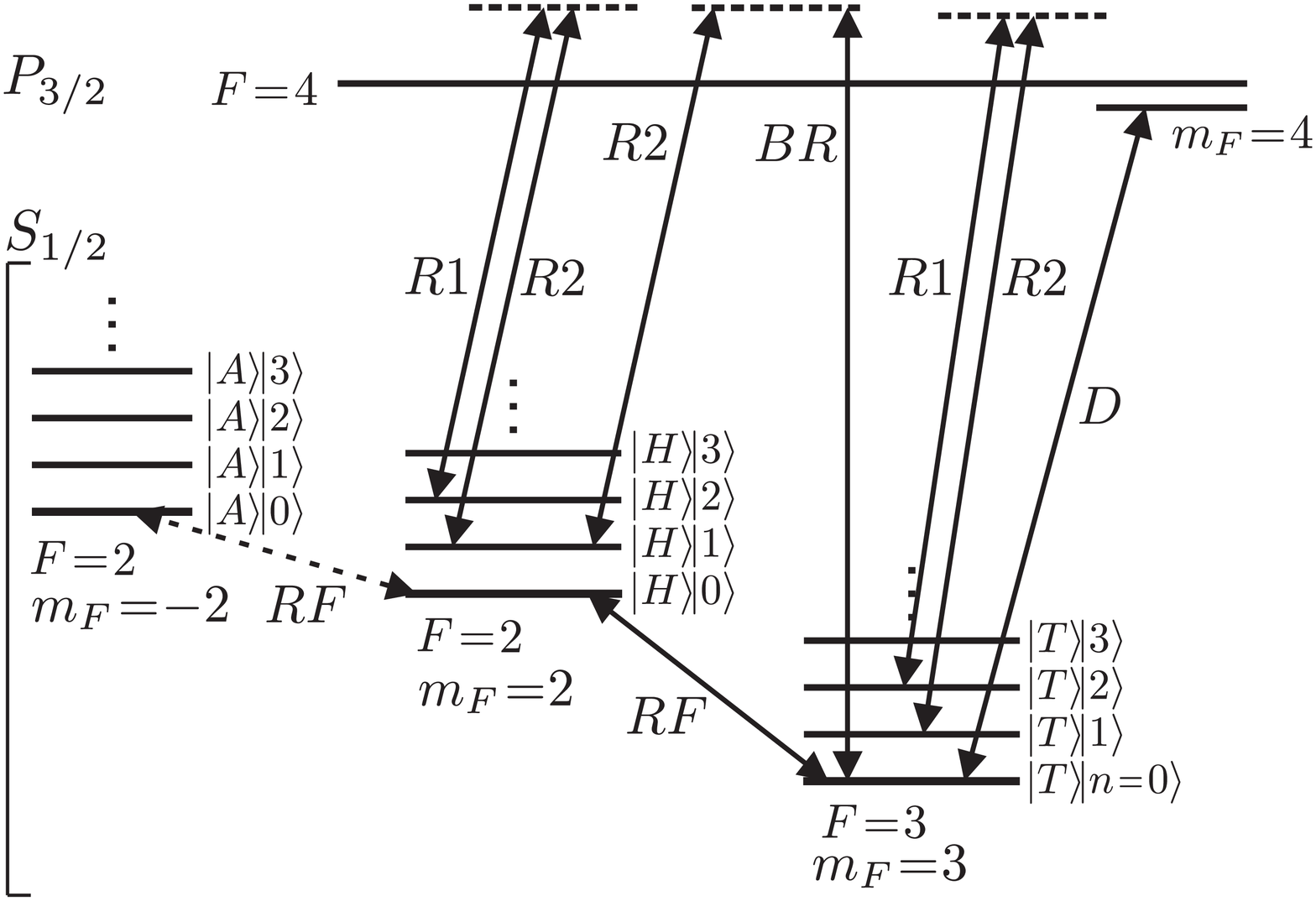}
\caption{Schematic of the relevant electronic and some of the lowest motional energy levels (not to scale) of $^{25}Mg^+$ and the transitions used for the QW experiment. A magnetic field, pointing into the direction of the laser beam R$1$, provides a Zeeman splitting of the hyperfine levels and the quantization axis of the system. The coin states are encoded in $\ket{^2S_{1/2}, F\text{=3}, m_F\text{=3}} = \ket{T}$ and $\ket{^2S_{1/2}, F\text{=2}, m_F\text{=2}} = \ket{H}$. Laser $D$, driving a closed cycling transition, in good approximation independent of the motional level $\ket{n}$, is used for optically pumping the electronic state into $\ket{T}$, for Doppler cooling of all motional modes \cite{Wineland1979} and readout of the internal state (See sect. \ref{detection}) \cite{Wineland1998}. The pair of laser beams $BR$ and $R2$ is used to drive a two-photon stimulated Raman transition on the red sideband of the coin state transition for sideband cooling of the axiaconsequences of leaving the scope of the approximations originally
used, such as the Lamb--Dicke regime.l motional direction \cite{Monroe1995} and on the blue sideband (BSB, shown here) for the readout of the motional state (See sect. \ref{detection}). The laser beams $R1$ and $R2$ drive a two-photon stimulated Raman transition providing the coin-state dependent optical dipole force (Figure \ref{dipolkraftschema}b). Additionally we apply a RF to drive coherent transitions between electronic states, independent of the motional state (Figure \ref{dipolkraftschema}a). With the RF we implement the coin operation and the transition via several steps from $\ket{H}$ to $\ket{A}=\ket{^2S_{1/2}, F=-2, m_F=2}$ required for the readout of the motional state (See sect. \ref{detection}).}
\label{termschema}
\end{figure}

Ideally, we encode the positions of the QW into coherent motional
states $\ket{\alpha_k}$ of the ion's axial harmonic motion. We
manipulate the motion by implementing the shift operator $S$
\eqref{stepop} via the application of a coin-state dependent optical
dipole force (See figure \ref{dipolkraftschema}b). In the following
we describe this method and its limitations.

\subsubsection{Experimental tools}
The initial motional state after sideband cooling is close to the
ground state $\ket{n=0}$ ($\overline{n}<0.03$). We apply a
two-photon stimulated Raman transition between the coin states by
applying two laser beams ($R1$, $R2$) (Figures
\ref{dipolkraftschema}b and \ref{termschema}), at a detuning of
$\Delta=2\pi\cdotp80 \text{ GHz}$ from the $P_{3/2}$ state manifold
and a fixed phase relation \cite{Wineland1998}. The frequency
difference between $R1$ and $R2$ amounts to
\begin{equation}
\omega_L = \omega_1 - \omega_2 = \omega_z - \delta,
\end{equation}
with $\delta = 2\pi \cdotp 100 \text{ kHz}$. The effective wave vector is $\bi{k_1}-\bi{k_2}=k\bi{e_z}$, pointing into the axial direction (Figure \ref{dipolkraftschema}b). This allows for two-photon stimulated Raman transitions $\ket{T}\ket{n}\leftrightarrow\ket{T}\ket{n+1}$ and $\ket{H}\ket{n}\leftrightarrow\ket{H}\ket{n+1}$ ($\forall\: n$). In a simplified picture the two laser beams provide a walking standing wave causing a state dependent ac-Stark shift. This yields a coin-state dependent force ($F_T$, $F_H$), proportional to the spacial gradient of the walking wave and oscillating with frequency $\omega_L$. The ratio of the forces acting on the coin states amounts to $F_H/F_T\approx -2/3$. The polarizations and intensities of the laser beams are adjusted such that the time-averaged ac-Stark shift for pulse durations $T\gg 0.5 \text{ }\mu\text{s}$ is negligible \cite{Wineland2003}. Thus the application of the dipole force does not change the relation between the relative phase of the coin states and the phase of the RF, which implements the coin operator. The effective wavelength of the walking wave amounts to $\lambda \approx 200 \text{ nm}$. With the width of the axial ground-state wave function of $z_0 \approx 10 \text{ nm}$ this results in a Lamb--Dicke parameter of $\eta = z_0\cdotp 2\pi/\lambda = 0.31$ \cite{Wineland1998}.

\subsubsection{Description of the dynamics}\label{dynamicsdescription}
We consider the following Hamiltonian describing a two-level system
coupled to a harmonic oscillator and interacting with a classical
light field \cite{Wineland1998}
\begin{equation}
\begin{split}
\mathcal{H} =& \mathcal{H}_{coin} + \mathcal{H}_{motion} + \mathcal{H}_{interaction}\\
=& \frac{\hbar}{2}\omega_{coin}\sigma_z + \hbar \omega_z (a^{\dagger}a + \frac{1}{2})+\hbar\underline{\Omega}_D\cos(k\cdotp z-\omega_Lt+\phi_0)\\
 =& \frac{\hbar}{2}\omega_{coin}\sigma_z + \hbar \omega_z (a^{\dagger}a + \frac{1}{2})+ \frac{\hbar}{2}\underline{\Omega}_D \left(e^{i(\eta (a + a^{\dagger})-\omega_Lt+\phi_0)}+h.c.\right).
\end{split}
\end{equation}
where
$\underline{\Omega}_D=\Omega_D\left(\ket{T}\bra{T}-\frac{2}{3}\ket{H}\bra{H}\right)$
with $\Omega_D$ the coupling factor and $\sigma_z$ the Pauli
$z$-matrix. The dynamics have been investigated for many
applications in QIP \cite{Leibfried2003a, Monroe1996} (in the LDA,
see below), for the simulation of nonlinear optics
\cite{Wallentowitz1997} and in the context of mesoscopic
entanglement \cite{McDonnell2007, Poschinger2010}. In the
interaction picture, with the free Hamiltonain being
$\mathcal{H}_{coin} + \mathcal{H}_{motion}$, the interaction
Hamiltonian can be written as
\begin{equation}\label{dipolkraft_allgemein}
\begin{split}
\mathcal{H}_I(t)  =&\frac{\hbar}{2} \underline{\Omega}_D \otimes \sum^{\infty}_{m=0} \sum^{\infty}_{n=0} \quad \ket{m}\bra{m}   e^{i \eta (a + a^{\dagger})}    \ket{n}\bra{n}\\
&\times\Bigl(e^{i((m-n) \omega_z - \omega_L ) t+i\phi_0} + (-1)^{\lvert m-n \rvert} e^{i((m-n) \omega_z + \omega_L ) t-i\phi_0} \Bigr).
\end{split}
\end{equation}
We apply the dipole force with a small detuning of
$\delta=\omega_z-\omega_L=2\pi\cdotp 100 \text{ kHz}$, such that the
terms corresponding to first-sideband transitions,
$\ket{n}\leftrightarrow\ket{n+1}$, rotate slowest and thus dominate.
However, we also consider contributions up to the third sideband,
$\lvert m-n\rvert=3$, an approximation we refer to as 3SB (Figures
\ref{omegas}, \ref{schritte} and \ref{scannah}).

When considering only the slowest rotating terms, i.e. applying the
usual rotating-wave approximation (RWA), the interaction Hamiltonian
is reduced to:
\begin{equation}\label{dipolkraft_rwa}
\begin{split}
\mathcal{H}^{RWA}_I(t) =& \frac{\hbar}{2} \underline{\Omega}_D \otimes \sum^{\infty}_{n=0} \Bigl( \bra{n+1} e^{i \eta (a + a^{\dagger})} \ket{n} e^{i( \delta t+\phi_0)} \ket{n+1}\bra{n} \\
&- \bra{n} e^{i \eta (a + a^{\dagger})} \ket{n+1} e^{-i (\delta t+\phi_0)} \ket{n}\bra{n+1} \Bigr).
\end{split}
\end{equation}

For states with $k\sqrt{\braket{z^2}}=\eta\sqrt{\braket{(a+a^\dagger)^2}}\ll 1$, we can approximate $\bra{n+1}e^{i\eta(a+a^\dagger)}\ket{n} \approx i\eta\sqrt{n+1}$ \cite{Leibfried2003}. That is, the potential
providing the dipole force changes linearly over the extension of the wave function. We can then simplify the interaction Hamiltonian to
\begin{equation}\label{dipolkraft_lda}
\begin{split}
\mathcal{H}^{LD}_I(t) = \frac{\hbar}{2} \underline{\Omega}_D \otimes i \eta (a^{\dagger}e^{i(\delta t+\phi_0)} - ae^{-i(\delta t+\phi_0)}).
\end{split}
\end{equation}
This is the Lamb--Dicke approximation (LDA) \cite{Wineland1998} or
the linear approximation, respectively. In the following, we set $\phi_0=0$ and $\hbar=1$.\\

\begin{figure}
\centering
\includegraphics[width=9.5cm]{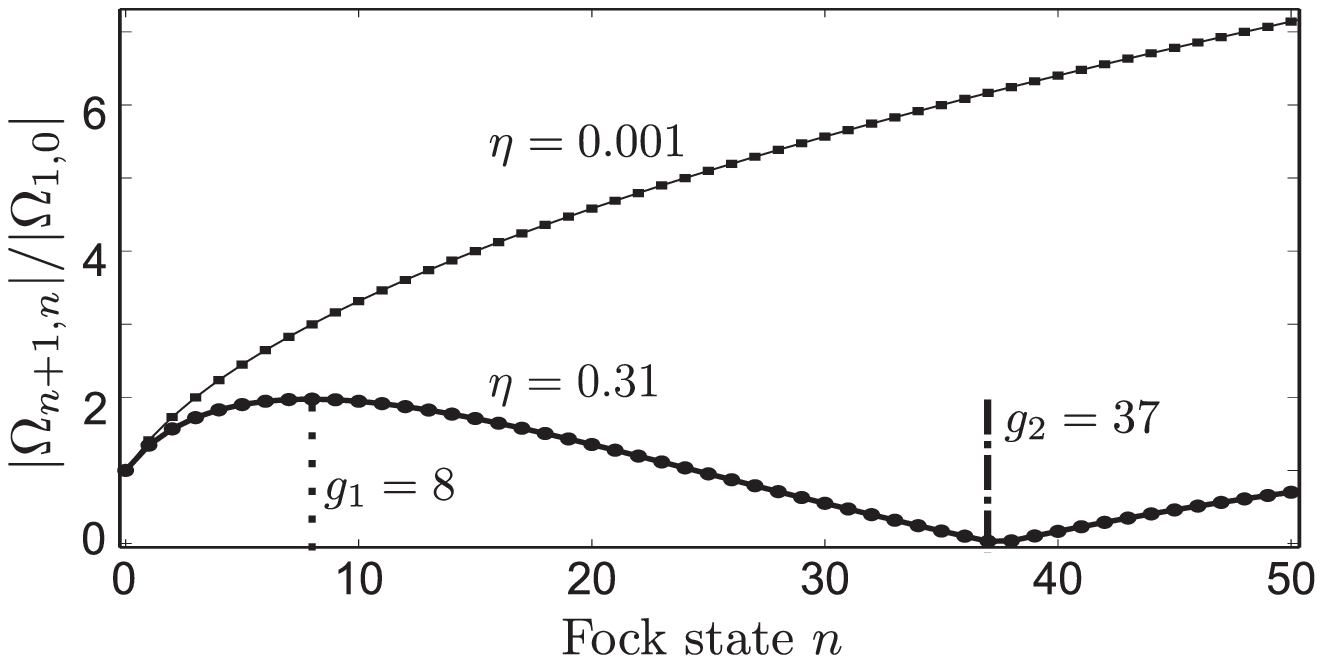}
\caption{
Normalized Fock-state transition rates $\lvert\Omega_{n+1,n}\rvert /\lvert\Omega_{1,0}\rvert=\lvert\bra{n+1}\exp\bigl(i\eta(a+a^{\dagger})\bigr)\ket{n}\rvert$ \eqref{dipolkraft_rwa} for a Lamb--Dicke parameter $\eta=0.31$ (experimental value) and $\eta=0.001$ (deeply within the LDR). As long as $\lvert\Omega_{n+1,n}\rvert /\lvert\Omega_{1,0}\rvert\approx\eta\sqrt{n}$, the force is constant over the ion's motional extension. The time evolution is then described by a displacement operator \eqref{ldatimeevo}, preserving the shape of coherent states. As the motional amplitude $\lvert\alpha\rvert=\sqrt{\braket{n}}$ increases, the force starts to remarkably change over the oscillating range of the wave function leading to motional squeezing.
The Fock state at which the matrix element is maximal, i.e. $n=8\equiv g_1$ for $\eta=0.31$, can be considered as the threshold above which severe motional squeezing of coherent states starts. The motional amplitude does not increase by the application of a continuous detuned force (See figure \ref{omegas}).
Motional states up to $g_2$ for $\eta=0.31$ can be created - with an increasing amount of squeezing - by applying the dipole force resonantly (Figure \ref{resonant}) or by step-wise excitation (Figure \ref{schritte}).
} \label{matrixelemente}
\end{figure}

Within the LDA, the time evolution operator is given by \cite{Carruthers1965}
\begin{equation}\label{ldatimeevo}
U (t) = \ket{T}\bra{T}\otimes D\left(\alpha(t)\right)\cdotp e^{i \Phi\left(\alpha(t), t\right)}+\ket{H}\bra{H}\otimes D\left(-\frac{2}{3}\alpha(t)\right)\cdotp e^{i \Phi\left(-\frac{2}{3}\alpha(t), t\right)},
\end{equation}
which is a displacement operator  $D\left(\alpha(t)\right)$ with a phase factor, where the phase amounts to
\begin{equation}\label{phase}
\Phi \left(\alpha(t), t\right) = \text{Im}\left( \int^t_0 d\tau \: \alpha^*(\tau) \frac{d\alpha(\tau)}{d\tau}\right).
\end{equation}
The factor $2/3$ in the displacement operator for the Heads-part
results from the difference of the state-dependent dipole force,
$F_H/F_T=-2/3$ (See figure \ref{dipolkraftschema}). The complex
parameter appearing in the displacement operator and in the phase
amounts to
\begin{equation}\label{alphacircle}
\alpha(t)=\frac{\eta\Omega_D}{2}\cdotp\int^t_0e^{i\delta t}dt = -i\frac{\eta\Omega_D}{2\delta} \cdotp\left(e^{i\delta t}-1\right)
\end{equation}
and corresponds to a circular trajectory in a co-rotating phase space, given by the interaction picture as
\begin{equation}\label{pr_def}
\begin{pmatrix}\text{Re}(\alpha(t))\\ \text{Im}(\alpha(t))\end{pmatrix} = \begin{pmatrix}\cos(\omega_z t) & \sin(\omega_z t) \\ -\sin(\omega_z t) & \cos(\omega_z t) \end{pmatrix} \begin{pmatrix} \frac{1}{2z_0}\braket{z}(t) \\ z_0 \braket{p}(t) \end{pmatrix},
\end{equation}
with $\braket{z}(t)$ and $\braket{p}(t)$ the expectation values of position and momentum.

For each coin state, the motional wave function is coherently
displaced along a circular trajectory in the co-rotating phase space
(Figure \ref{omegas}). The circular shape of the trajectory is
caused by the dipole force being applied with a detuning $\delta$
relative to the oscillator frequency. Thus the relative phase
between dipole force and oscillation of the ion evolves in time as
$\phi_D(t)=\delta\cdotp t$. After a duration of $T_\pi=\pi/\delta$
of driving the motional state and increasing its amplitude, the
relative phase amounts to $\phi_D(T_\pi)=\pi$ and therefore the
dipole force starts to decelerate the oscillation of the ion. After
a duration of $T_{2\pi}=2\pi/\delta$ the coherent state returns to
its initial location in the co-rotating phase space. The total
acquired phase of the motional state, which equals the enclosed area
of the trajectory \cite{Leibfried2003a}, amounts to
\begin{equation}
\Phi_T=\pi\left(\frac{\eta\Omega_D}{2\delta}\right)^2
\label{phasefactor}
\end{equation}
for $\ket{T}$ and $\Phi_H=(4/9)\cdotp\Phi_T$ for $\ket{H}$.

\begin{figure}
\centering
\includegraphics[width=8cm]{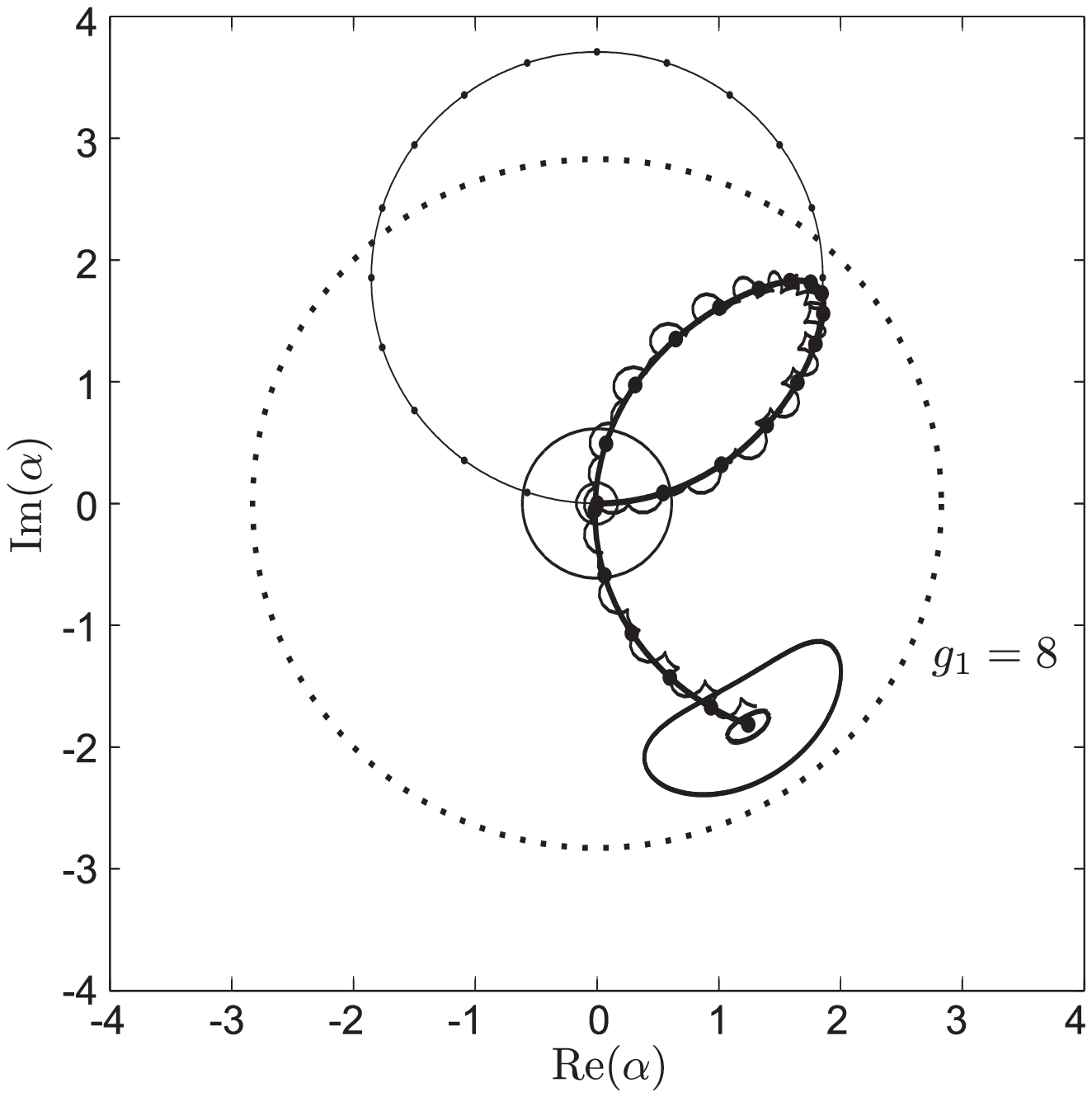}
\caption{Numerical simulation of the ion's trajectory in co-rotating phase space \eqref{pr_def} driven by the dipole force for the three relevant approximations 3SB \eqref{dipolkraft_allgemein}, RWA \eqref{dipolkraft_rwa} and LDA \eqref{dipolkraft_lda}.
The initial state is at the origin ($\ket{\alpha_0=0}$). The thin concentric lines represent contours of its Wigner function $W$ (at $W_>=0.6$ and $W_<=0.3$).
The bold dotted line represents $g_1$ (Figure \ref{matrixelemente}).
The thin circular trajectory with dots represents the result of the simulation within the LDA. The dots on the trajectory depict the positions after $t=0, 0.5, ... 10 \text{ \textmu s}$. The final state, reached after $T_{2\pi}=2\pi/\delta=10 \text{ \textmu s}$, equals the initial one, up to a phase factor. The bold trajectory represents the result within the RWA, taking nonlinearities of the dipole force into account (Figure \ref{matrixelemente}). The dots on the trajectory again depict the position at the times $t=0, 0.5, ... 10 \text{ \textmu s}$. Starting from the origin, the trajectory is identical to the one within the LDA. In the vicinity of $g_1$ the trajectories start to deviate. The acceleration of the ion ceases at a certain amplitude, the state gets squeezed and then returns to the origin after a duration shorter than $T_{2\pi}$. The spiraling trajectory, which follows the one within the RWA, represents the results within 3SB. Here terms of higher frequencies in the Hamiltonian are taken into account. The final Wigner function is almost identical to the one in the RWA and therefore not shown.
Parameters: $\Omega_D=2\pi \cdot 1.2 \text{ MHz}$, $\omega_L = 2\pi\cdot 2.03 \text{ MHz}, \omega_z = 2\pi\cdot 2.13 \text{ MHz}$, $\eta=0.31$, $t \in [0, 10] \text{ \textmu s}$.}
\label{omegas}
\end{figure}

The nonlinearity of the potential causing spatial variations of the
dipole force can be described by the absolute values of the
transition matrix elements, i.e. $\Omega_{n+1,n} =
\bra{n+1}\exp\bigl(i \eta (a + a^{\dagger})\bigr)\ket{n}$ (Figure
\ref{matrixelemente}). For small Fock state numbers $n$ they remain
close to the approximative results within the LDA (where the
potential is linear in $z$). Close to $n=8\equiv g_1$ (for
$\eta=0.31$) they start to significantly deviate from that
approximation. In particular, the transition matrix elements feature
a maximum value at $g_1$. The excitation of motion via the dipole
force ceases at $n=37\equiv g_2$, due to
$\lvert\Omega_{38,37}\rvert\approx 0$. It therefore represents an
upper bound for the motional excitation with a dipole force applied
close to resonance. (Applying the dipole force with a frequency
$\omega_L=2\cdotp\omega_z$ would allow populating higher Fock
states, but since the overall time evolution is then described by a
squeezing operator, it cannot be used for the implementation of a QW
based on coherent displacements, following reference
\cite{Travaglione2002}.)

Figure \ref{omegas} presents the results of our numerical
simulations of the time evolution, comparing the three different
approximations 3SB, RWA and LDA. Starting in the motional ground
state, the trajectory in the co-rotating phase space first follows
the circular evolution, as long as the amplitude remains small, i.e.
$\braket{n}<g_1$. This regime can be well described by the LDA (with
the driving potential being linear in $z$). As the amplitude of the
motional state approaches $g_1$, the driving potential becomes
sufficiently nonlinear to severely affect the subsequent evolution.
The amount of displacement per time interval is substantially
reduced, as the transition rates $\lvert\Omega_{n+1,n}\rvert$
decrease for $n\geq g_1$, such that the trajectory remains in the
vincinity of $g_1$. At this point motional squeezing occurs. The
probability distribution of the wave function in the Fock state
basis becomes narrower than Poissonian, which results in a squeezed
shape of the corresponding Wigner function. The relative phase
between the dipole force and the oscillation of the ion changes
faster than in the linear case. Thus the squeezed wavefunction
reaches the origin of the phase space after a time significantly
shorter than $2\pi/\delta$. The dependence of the return time and
the amount of squeezing on the maximal motional amplitude severely
affect an implementation of a QW with position states outside the
$LDR$, following the scheme described in reference
\cite{Travaglione2002}. However, in section \ref{outlook} we propose
an alternative protocol for the implementation of the shift operator
that circumvents this restriction. The faster rotating terms, which
are taken into account in the 3SB approximation, cause additional
modulations of the trajectory with low amplitudes and high
frequencies $(2\omega_z+\delta)$ and $(3\omega_z+\delta)$,
respectively.

\begin{figure}
\centering
\includegraphics[width=14cm]{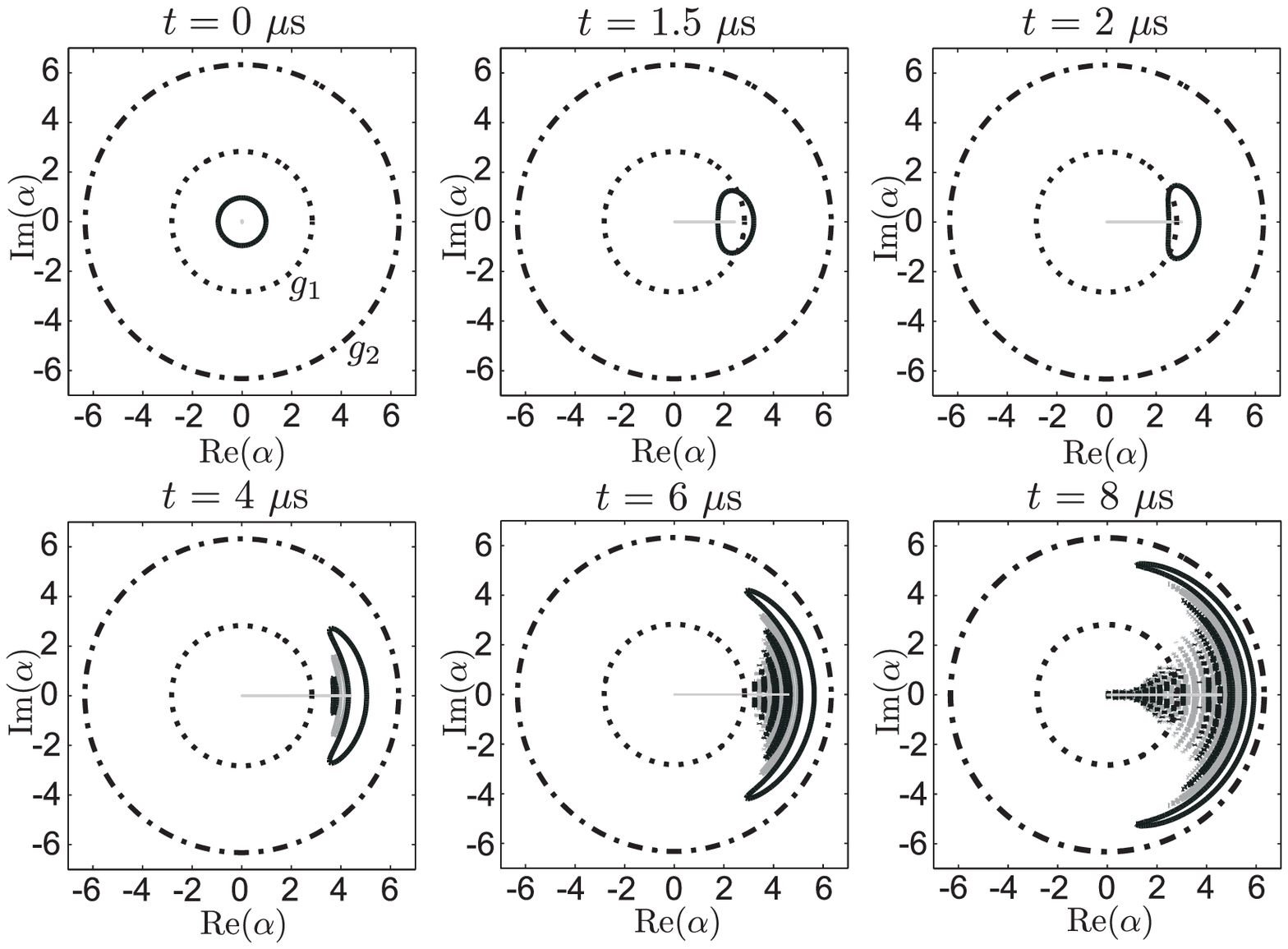}
\caption{Numerical simulation (3SB) of a resonant ($\delta=0$) excitation from the motional ground state, trajectory (grey horizontal line) and contours of the Wigner function $W$ (black: $W=0.1$, grey: $W=-0.1$) for increasing durations $t=\{0, 1.5, 2, 4, 6, 8\} \text{ \textmu s}$. Starting from the ground state, the dipole force continuously displaces the ion in good approximation to coherent states along the real axis up to $g_1$ (dotted circle). Further excitation comes along with severe motional squeezing. Amplitudes higher than $g_2$ are not considerably populated, due to the almost vanishing matrix element $\Omega_{37,38}$ at $g_2$ (dash-dotted circle) (Figure \ref{matrixelemente}). As a consequence, at $g_2$ the wave function gets reflected in the sense that the amplitude gets decreased again. The interference of the accelerated and the decelerated (reflected) part of the wave function is resembled in the Wigner function, which shows concentric lines of positive and negative value in the populated area of the rotating phase space. Parameters: $\Omega_D=2\pi \cdot 2.0 \text{ MHz}$, $\omega_L = \omega_z = 2\pi\cdot 2.0 \text{ MHz}$, $\eta=0.3$.
}
\label{resonant}
\end{figure}

Further motional excitation (up to $g_2$) can be realized either by
applying the dipole force resonantly (Figure \ref{resonant})
\cite{Wallentowitz1997}, or by the repeated off-resonant application
of a weak dipole force with duration $\pi/\delta$ and constant time
delays between the pulses (Figure \ref{schritte}). However, in
either way severe motional squeezing occurs as the amplitude
approaches and becomes larger than $g_1$.

\begin{figure}
\centering
\includegraphics[width=16cm]{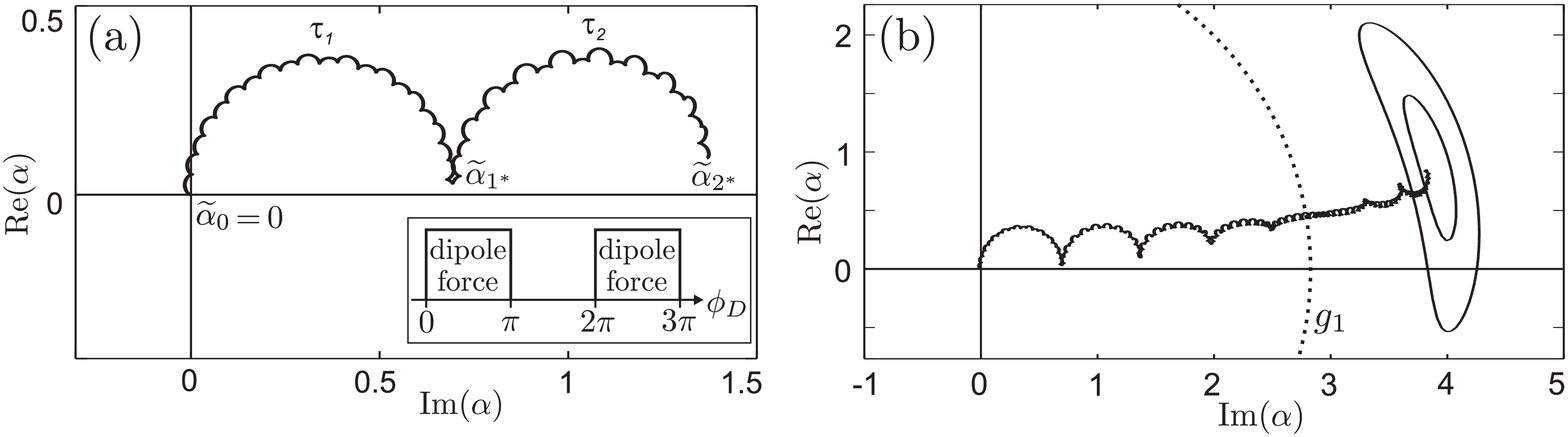}
\caption{Numerical simulation (3SB) of the ion's trajectory in the co-rotating phase space during a stepwise excitation. (a) The first dipole force pulse (inlay) displaces the ion from the ground state to $\posr{1^*}$ (which is not the position state $\posr{1}$, compare figure \ref{kombipuls}) along trajectory $\tau_1$. During the pulse the relative phase between dipole force and oscillator motion increases to $\phi_D=\pi$. The phase difference further increases after the dipole force is switched off (Sect. \ref{stepopimplement}). After the duration $t=\pi/\delta$, it amounts to $\phi_D=2\pi$. Applying now a second dipole force pulse (inlay) displaces the ion to $\posr{2^*}$ along trajectory $\tau_2$.
(b) Stepwise excitation as described above, with eight steps, passing the threshold $g_1$. As the state passes $g_1$, the trajectory during each dipole force pulse follows the opposite direction of rotation (counter-clockwise). In this regime (in particular close to $g_2$) the time evolution can be approximated by a displacement operator as in \eqref{ldatimeevo} parametrized with $\alpha(-t)$ \cite{Wallentowitz1997}. The contours of the final Wigner function show the strong motional squeezing. In addition, due to the dependency of the return time on the amplitude (Cf. Figure \ref{omegas}) the trajectory is on average not oriented along the horizontal axis. This is also the case in the protocol of the three-step QW (Figure \ref{pulsschema}), but does not affect its performance. Parameters: $\Omega_D=2\pi \cdot 0.4 \text{ MHz}$, $\omega_z = 2\pi\cdot 2.0 \text{ MHz}$, $\delta=2\pi \cdot 100 \text{kHz}$, $\eta=0.3$.}
\label{schritte}
\end{figure}

\subsubsection{Implementation of the shift operator}\label{stepopimplement}
We implement several shifts into a certain direction in the
co-rotating phase space by a synchronized application of dipole
force pulses. Switching the dipole force is realized by
acousto-optical modulators, refracting the laser beams $R1$ and $R2$
into the Paul trap \cite{Wineland1998}. As the laser is continuously
on during the whole experiment, the phase relation between the
dipole force and the motion of the ion continuously evolves in time
as $\phi_D(t)=\delta\cdotp t$, even when the optical dipole force is
switched off. Therefore, after the first pulse of the dipole force,
the relative phase $\phi_D(t)$ and thus the direction of the
displacement caused by the following dipole force pulse depends on
the intermitted delay. We apply pulses of the duration
$T^{QW}_D\approx\pi/\delta=5\text{ \textmu s}$ and mutual delays
that concatenate the displacements along a line in the co-rotating
phase space (Figure \ref{schritte}).

\begin{figure}
\centering
\includegraphics[width=12cm]{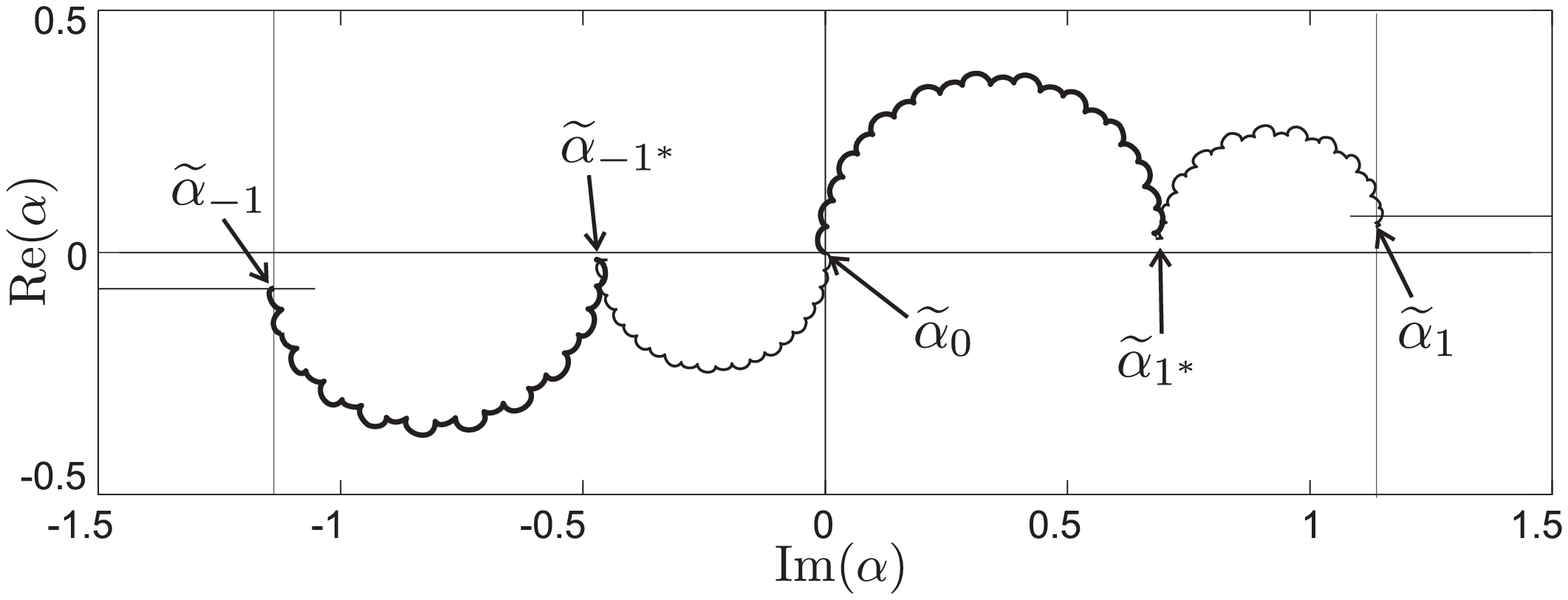}
\caption{Numerical simulation (3SB) of the trajectories related to the ion's coin states $\ket{T}$ (bold trajectory) and $\ket{H}$ (thin trajectory) during step 1 of the QW (Figure \ref{pulsschema}). After the coin operation the ion's state is $\ket{\psi}=(\ket{T}+\ket{H})\posr{0}$ (where $\posr{k}$ denotes the possibly slightly squeezed version of $\posi{k}$). The shift operation $S$ is implemented by a $combined$ $pulse$. The first dipole force pulse displaces the motional state related to $\ket{T}$ ($\ket{H}$) to $\posr{1^*}$ ($\posr{-1^*}$). As the forces are of different amplitudes, the two different trajectories lead to different phase factors, related to $\Phi_T$, $\Phi_H$ \eqref{phasefactor}. Subsequently the coin states are exchanged via an $R(\pi,0)$-pulse without affecting the motional states and after a specific waiting duration (Figure \ref{pulsschema}) a second dipole force pulse is applied, displacing the motional states to $\posr{-1}$ ($\posr{1}$). A second $R(\pi,0)$-pulse exchanges the coin states again, such that the resulting state of the ion is $\ket{\psi_1}=\exp\bigl(i(\Phi_T+\Phi_H)\bigr)\cdotp (\ket{T}\posr{1}+\ket{H}\posr{-1})$, accumulating only a global phase during the shift operation. The combined-pulse scheme further provides equal step distances in both directions.
Parameters: $\Omega_D=2\pi \cdot 0.24 \text{ MHz}$, $\omega_z = 2\pi\cdot 2.13 \text{ MHz}$, $\delta=2\pi \cdot 100 \text{kHz}$, $\eta=0.31$.}
\label{kombipuls}
\end{figure}

\begin{figure}
\centering
\includegraphics[width=13cm]{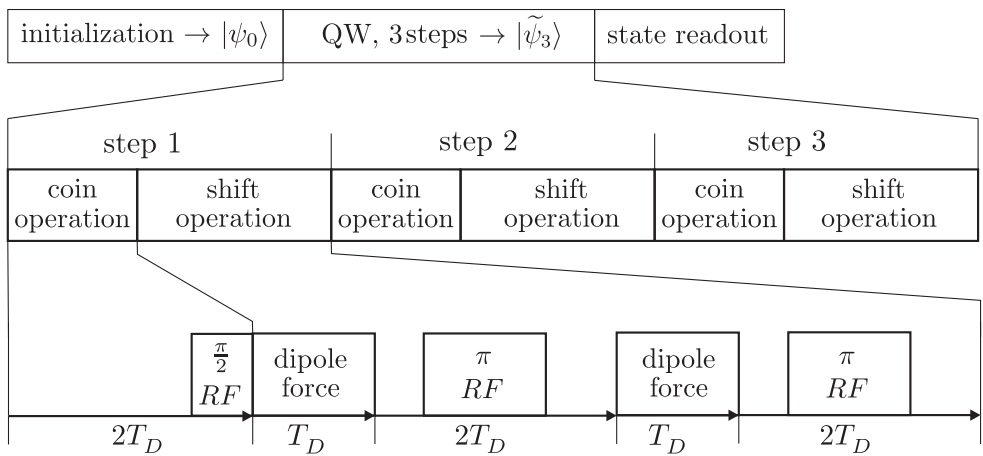}
\caption{Schematic of the pulse sequence for the implementation of three steps of an asymmetric QW. The first line (top) illustrates the overall pulse sequence for one experimental cycle, in chronological order from left to right. We initialize the ion in the state $\ket{\psi_0}=\ket{T}\ket{\alpha_0}$. Then we apply the pulse sequence for the QW, creating the state $\ket{\widetilde{\psi}_3}$ \eqref{realstate}, and finally a pulse sequence to readout the coin or motional state (Sect. \ref{detection}). The pulse sequence for the QW consists of three subsequent applications of coin $C$ (Figure \ref{dipolkraftschema}) and shift $S$ (Figure \ref{kombipuls}) operations, implementing the three steps of the QW as illustrated in the second line. The third line depicts the pulses implementing one coin and one shift operation. The phase $\phi$ (not depicted) is equal for every RF pulse (Sect. \ref{coinimplementation}). The sequence of all RF pulses incorporates a spin-echo scheme, reducing dephasing of the coin states. A symmetric QW can be implemented with the same scheme, but with the phase of the initial RF pulse differing by $\pi/2$ from the (equal) phases of all other RF pulses. The timing of the pulses, in particular for the dipole force, is given by the parameter $T_D$, as illustrated in the bottom. It must be chosen such that the overlap of the interfering motional states (Figure \ref{schrittidee}) is maximal. The theoretical optimal value within the LDR would be $T_D=\pi/\delta$. We determine the optimal value of $T_D$, denoted as $T^{QW}_D\approx\pi/\delta$, experimentally to account for the nonlinearity of the dipole force as well as for experimental imperfections (Figure \ref{scannah}).
}
\label{pulsschema}
\end{figure}

However, the motional states corresponding to $\ket{H}$ and
$\ket{T}$ acquire different phase factors $\Phi_H$, $\Phi_T$
\eqref{phasefactor} during each displacement. To compensate these
coin-state dependent phases, we implement the shift operation of the
QW as a $combined$ $pulse$, which consists of two dipole force
pulses, each followed by an $R(\pi,0)$-pulse (Figure
\ref{kombipuls}). In this scheme each coin state acquires the sum of
the phase factors, $\Phi_T+\Phi_H$, which therefore turns into a
global phase factor not affecting the QW. A schematic of the overall
pulse sequence for the QW is depicted in figure (\ref{pulsschema}).

For the implementation of the QW the following is crucial. After a
step fulfilling the operation
$\ket{T}\posr{k}\rightarrow\ket{T}\posr{k+1}$, the subsequent coin
toss and shift operation have to ensure the operation
$\ket{H}\posr{k+1}\rightarrow\ket{H}\posr{k}$, i.e. the motional
state $\posr{k+1}$ must be transfered back to the previous state
$\posr{k}$. This must be fulfilled for all $k$ simultaneously (See
figure \ref{schrittidee}). In order to reach the state $\posr{k}$,
the duration $T_D$ and detuning $\delta$ of the dipole force have to
be adjusted properly (In the LDR to $T_D=\pi/\delta$), implementing
semi-circular trajectories in the co-rotating phase space. But since
the return time is reduced outside the LDR (Figure \ref{omegas}),
the shift operation implements the transition from $\posr{k+1}$ to
some other state $\posr{k_2}\neq\posr{k}$. The reduced overlap
$\braket{\widetilde{\alpha}_{k_2}|\widetilde{\alpha}_{k}}<1$ leads
to reduced interference during the succeeding coin toss (Figure
\ref{schrittidee}).

In principle the shift operator can alternatively be implemented by
the dipole force on resonance ($\delta=0$). However, in that case
small variations of $\omega_L$ and $\omega_z$ have a much stronger
influence than in the detuned case. This can be seen by comparing
the difference in displacement, i.e.
$\lvert\alpha(\delta,t)-\alpha(\delta+\epsilon,t)\rvert$, for a
fixed duration $t$ and a fixed difference in the detuning,
$\epsilon$, using \eqref{alphacircle} for different values of
$\delta$, where $\alpha(\delta,t)$ fulfills equation
\eqref{alphacircle} for the detuning $\delta$.

\subsection{State readout}\label{detection}
The state after three steps of the QW is
\begin{equation}\label{realstate}
\ket{\widetilde{\psi}_3}=\frac{1}{\sqrt{2}}\left(\ket{T}\ket{M_T}+\ket{H}\ket{M_H}\right),
\end{equation}
with $\ket{M_T}=\sum^3_{k=-3} c^T_k \posr{k}$ and
$\ket{M_H}=\sum^3_{k=-3} c^H_k \posr{k}$ (Cf. \eqref{psi3}). The
basics for the readout are state-of-the-art techniques in quantum
information processing with trapped ions \cite{Meekhof1996,
Wineland1998}.

We readout the coin state by driving the cycling transition
$\ket{T}\ket{n} \rightarrow \ket{2P_{3/2},F=4,m_F=4}\ket{n}$ (Figure
\ref{termschema}) for a duration of 20 \textmu s and detect
scattered photons with a photomultiplier. This transition is in good
approximation independent of the motional state. The average number
of detected photons is proportional to the probability
$P_T(\widetilde{\psi}_3)$, related to the coin state $\ket{T}$
\cite{Wineland1998}. The probability $P_H(\widetilde{\psi}_3)$ is
accessible via the application of a $R\left(\pi,0\right)$-pulse
before the coin-state detection.

To characterize the motional states of $\ket{\widetilde{\psi}_3}$ we
determine the position-state probabilities $\lvert c^T_k \rvert^2$
and $\lvert c^H_k\rvert^2$. To analyze $\ket{T}\ket{M_T}$, we
isolate the other part, $\ket{H}\ket{M_H}$, by transfering
$\ket{H}\ket{M_H} \rightarrow \ket{A}\ket{M_H}$ using appropriate RF
pulses (Figure \ref{termschema}). The part $\ket{A}\ket{M_H}$ of the
ion's state is not affected by the subsequent operations. We then
apply a two-photon stimulated Raman transition $\ket{T}\ket{n}
\leftrightarrow \ket{H}\ket{n+1}$ ($\forall \: n$) using the lasers
$BR$ and $R2$ (Figure \ref{termschema}), with a frequency difference
of $\omega_L = \omega_{coin}+\omega_z$ (BSB) for a variable duration
$t_{BSB}$ \cite{Wineland1998}. The corresponding Rabi frequency for
each $n$ is proportional to $\Omega_{n+1,n}$ (Figure
\ref{matrixelemente}). Finally, we apply the coin-state detection,
as described above. The average number of detected photons is
proportional to the probability \cite{Wineland1998}
\begin{equation}\label{bsbkurve}
P_T(t_{BSB}) = \frac{1}{2}\left(1+\sum^\infty_{n=0} \lvert a^T_n\rvert^2 \cdotp \cos(\Omega_{n+1,n}\cdot t_{BSB}) e^{-\gamma \cdotp t_{BSB}}\right),
\end{equation}
with $a^T_n=\braket{n|M_T}$ the coefficients of $\ket{M_T}$ in the
Fock state basis. The damping factor $\gamma$ accounts for
decohering effects, mainly of the BSB-operation \cite{Wineland1998}.
A discrete Fourier transform of the function $P_T(t_{BSB})$ allows
to access the Fock state probabilities $\lvert a^T_n\rvert^2 =
\lvert \sum_k c_k^T \braket{n| \widetilde{\alpha}_k}\rvert^2$.

\begin{figure}
\centering
\includegraphics[width=12cm]{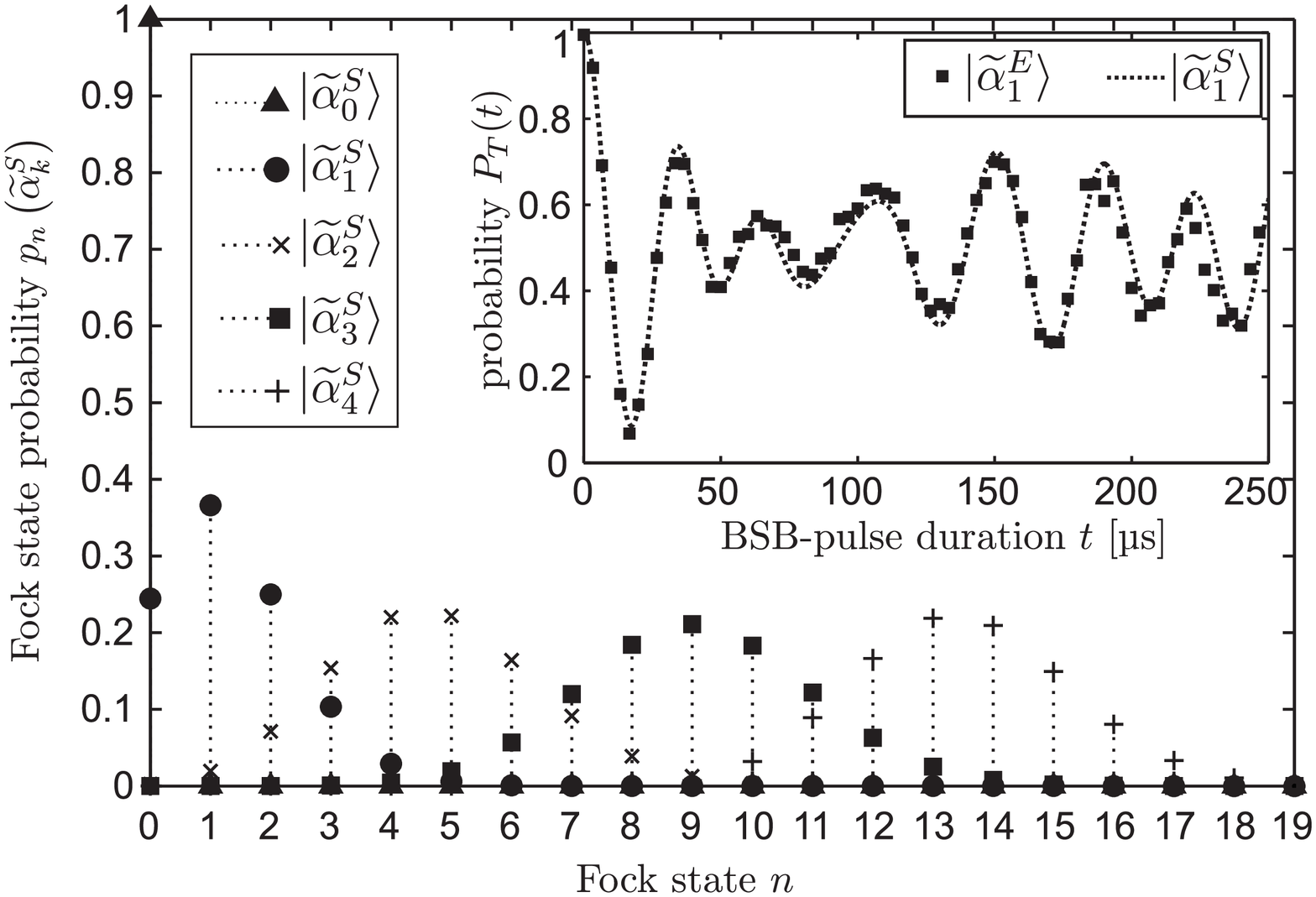}
\caption{Fock state probabilities $p_n\left(\widetilde{\alpha}^S_k\right)=\lvert\braket{n|\widetilde{\alpha}^S_k}\rvert^2$ of the simulated (3SB) position states $\poss{k}$ resembling the probabilitites $p_n(\widetilde{\alpha}^E_k)$ of the experimental position states $\pose{k}$.
Experimentally, we apply the shift operation $S$ (Figure \ref{kombipuls}) $k$ times (without coin operations) to the initialized ion and readout the motional state (Sect. \ref{detection}).
The inlay shows the coin state probability $P^E_T(t)$ (black squares) from the motional state readout of $\ket{T}\pose{1}$, as an example. Each experimental data point represents the average of 3000 realizations. A simulation (3SB) of this procedure is optimized such that the values $P^S_T(t)$ (dotted line, cf. \eqref{bsbkurve}) correspond to the experimental result for every position state $\poss{k}$, as shown for $\poss{1}$ in the inlay. The related expectation values of the position states in the Fock basis (main figure) are $\braket{n}^S_0=0$, $\braket{n}^S_1=1.33$, $\braket{n}^S_2=4.71$, $\braket{n}^S_3=9.08$, $\braket{n}^S_4=13.50$.
Their mutual overlaps amount to $\lvert\braket{\alpha^S_k|\alpha^S_{k+1}}\rvert^2\approx0.24$.
}
\label{positionen}
\end{figure}

We numerically simulate the QW and generate a corresponding function
$P^S_T(t_{BSB})$ (where the index $^S$ denotes the result of the
simulation), and optimize the parameters of the simulation in order
to fit $P^S_T(t_{BSB})$ onto the experimental data, $P^E_T(t_{BSB})$
(indicated by the index $^E$). The state
$\ket{\widetilde{\psi}_3^S}$, generated by the simulation, resp. the
Fock-state probabilities of its motional parts, are fed into the
algorithm for the discrete Fourier transform of $P^E_T(t_{BSB})$ to
improve the convergence of the results. The Fock state probabilities
$p_n\left(\widetilde{\alpha}_k\right)=\lvert\braket{n|\widetilde{\alpha}_k}\rvert^2$
of the position states $\posr{k}$ are determined separately (Figure
\ref{positionen}), using the same method.

To identify the position state probabilities $\lvert c^T_k\rvert^2$ from the Fock state probabilities of $\ket{M_T}$, in particular to distinguish between the coefficients related to $p_n\left(\widetilde{\alpha}_k\right)$ and $p_n\left(\widetilde{\alpha}_{-k}\right)$, we additionally apply the motional-state readout to the shifted states $\ket{\widetilde{\psi}^+_3}= S\ket{\widetilde{\psi}_3}$ and $\ket{\widetilde{\psi}^-_3}= S^{-1}\ket{\widetilde{\psi}_3}$ (See \eqref{stepop} and figure \ref{qw_verteilungen}), where $S^{-1}$ is (up to a global phase) implemented by a proper timing of the corresponding dipole force pulses.

\section{Experimental procedure}\label{realization}
\subsection{Calibration of the step size $\lvert\Delta\alpha\rvert$}\label{stepsizecalibration}
Starting from the initial state $\ket{\psi_0}$ we apply 0 to 4 shift
operations $S$ with a dipole force duration of
$T_D\approx\frac{\pi}{\delta}$, exciting the motion of the ion to
one of the position states $\posr{k}$. Then we readout the motional
state to determine its Fock state probabilities
$p_n\left(\widetilde{\alpha}_k\right)$ and the expectation of the
number operator $\braket{n}\equiv \lvert
\widetilde{\alpha}_k\rvert^2$. Small deviations of the dipole force
duration (See next subsection) do not influence the probabilities
significantly. We adjust the amplitude of the dipole force by
adjusting the corresponding laser beam intensities to approximately
meet the conditions $\lvert\Delta\alpha\rvert \geq 1$ and $\lvert
\widetilde{\alpha}_3\rvert
^2\equiv\bra{\widetilde{\alpha}_3}n\ket{\widetilde{\alpha}_3} \leq
9$ (three well distinguishable position states within or close to
the LDR). The Fock state expectation values of the position states
amount to
$\braket{n}_0=\bra{\widetilde{\alpha}_0}n\ket{\widetilde{\alpha}_0}=0$,
$\braket{n}_1=1.33$, $\braket{n}_2=4.71$, $\braket{n}_3=9.08$,
$\braket{n}_4=13.50$. The outer position states $\posr{\pm3}$ and
$\posr{\pm4}$ are not within the LDR. The step sizes therefore
differ, reaching from
$\lvert\widetilde{\alpha}_1\rvert-\lvert\widetilde{\alpha}_0\rvert=1.15$
to
$\lvert\widetilde{\alpha}_4\rvert-\lvert\widetilde{\alpha}_3\rvert=0.66$.
However, due to the motional squeezing the overlaps of all
neighbouring states amount to
$\lvert\braket{\widetilde{\alpha}_k|\widetilde{\alpha}_{k+1}}\rvert^2\approx0.24<1/e$,
which corresponds to the overlap of coherent states with a step size
of $\lvert\Delta\alpha\rvert>1$ as assumed in the theory (See
section \ref{theory}). With $\delta = 2\pi\cdotp 100 \text{ kHz}$
and $\eta=0.31$ the corresponding coupling strength of the dipole
force amounts to $\Omega_D = 2\pi\cdotp 0.24 \text{ MHz}$. The
amplitude of the corresponding dipole force amounts to $F_T =
(\hbar\eta\Omega_D)/(2z_0) = 2.54 \cdotp 10^{-21} \text{ N}$ inside
the LDR.

To estimate the amount of motional squeezing (not affecting the
fidelity of our results on the QW), we compute coherent states
$\posi{k}$ according to equation \eqref{alphadef} with
$\alpha_k\equiv\widetilde{\alpha}_k$. We then compute their overlaps
$F_k=\lvert\braket{\widetilde{\alpha}_k|\alpha_k}\rvert^2$, which
amount to $F_0=1.00$, $F_1=1.00$, $F_2=0.97$, $F_3=0.90$,
$F_4=0.78$.

\subsection{Calibration of the dipole force duration}
\begin{figure}
\centering
\includegraphics[width=12cm]{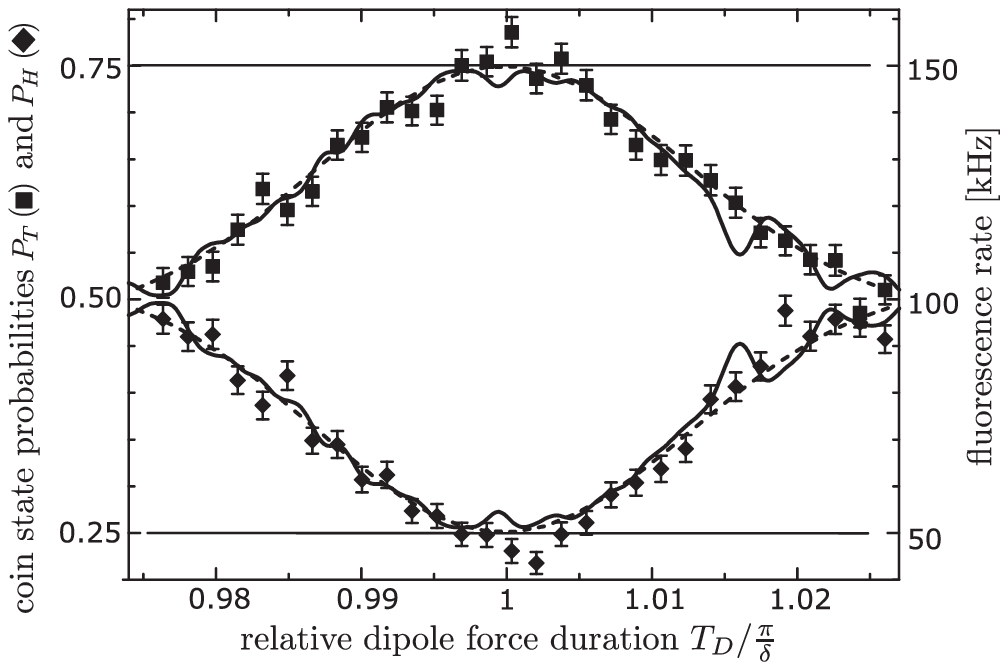}
\caption{Coin state probabilities in dependence of the relative dipole force pulse duration ($T_{D}/(\frac{\pi}{\delta})$) after the application of the QW pulse sequence (Figure \ref{pulsschema}) to the ion in state $\ket{\psi_0}$, and the corresponding numerical simulations (solid line: 3SB, dashed line: RWA). Each data point represents the average of 1500 realizations. The maximum ratio $P_H/P_T=1/3$ indicates the asymmetry due to interference (See figure \ref{schrittidee}). The corresponding dipole force duration, denoted as $T^{QW}_D$, is the optimal value to perform the QW. The precise value of $T^{QW}_D$ depends on the detuning $\delta$ which is prone to slow drifts of the conditions of the experimental setup (on a time scale of a few hours, much longer than an experiment) and can be estimated to the required precision (for each experiment) by this method. For other values of $T_{D}$ (i.e. $2 \%$ longer or shorter) the effect of the interference vanishes, since the shift operations of the pulse sequence do not lead to mutual overlaps of the related parts of the wave function.
In this experiment the QW pulse sequence (Figure \ref{pulsschema}) contains waiting durations of $4T_{D}$ instead of $2T_{D}$, which increases the sensitivity of the interference to $T_D$ and therefore allows for a more precise estimation of the optimal dipole force duration $T^{QW}_D$.
The 3SB simulation (solid line), in contrast to the RWA simulation (dashed line), contains a high-frequency modulation of the coin state probabilities due to additionally modulated overlap of the interfering states caused by the spiralling trajectories during the displacements (Figure \ref{omegas}). This is however not yet resolved in the experimental data.}
\label{scannah}
\end{figure}
\begin{figure}
\centering
\includegraphics[width=12cm]{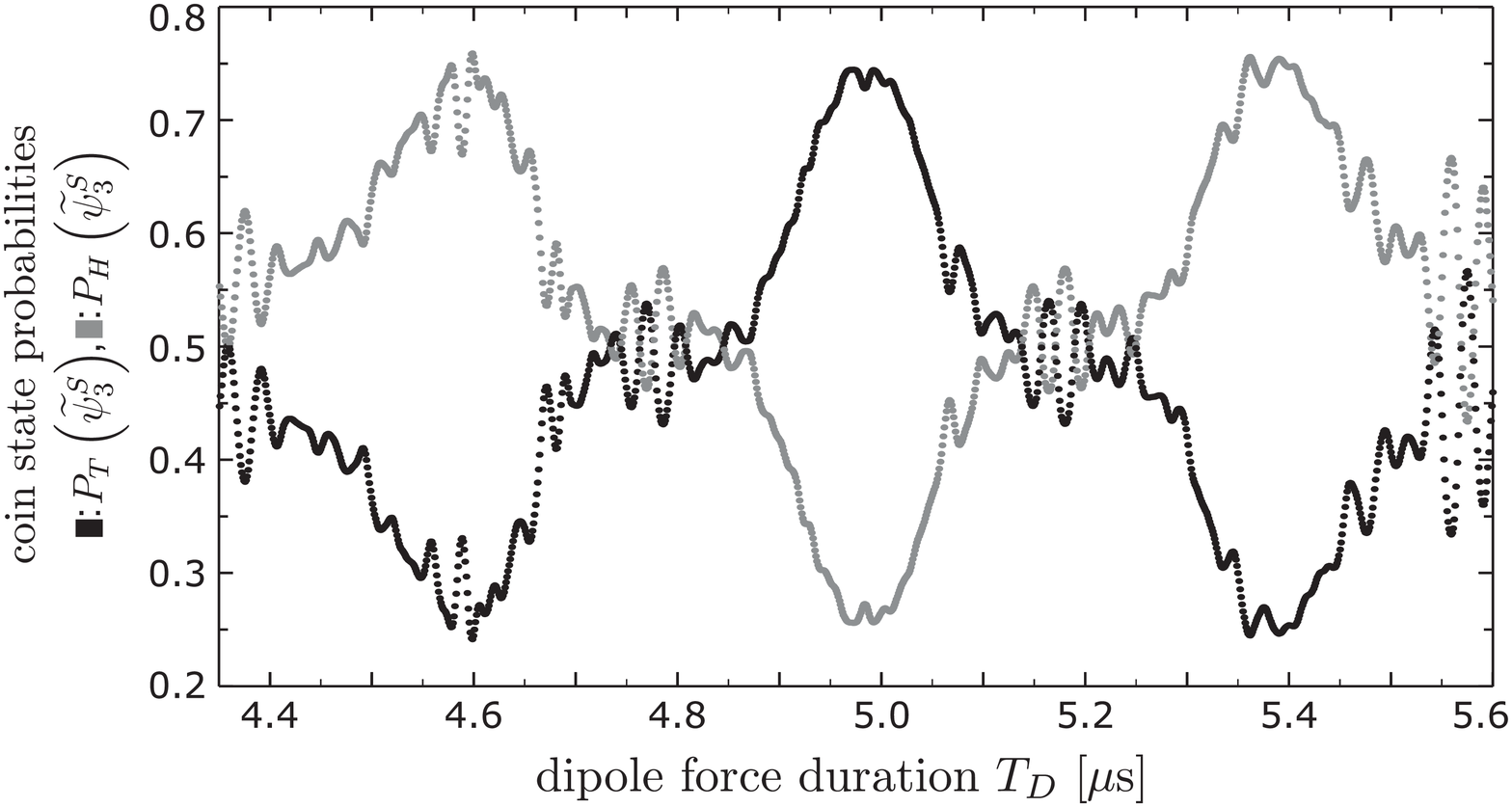}
\caption{Numerical simulation (3SB) as depicted in figure \ref{scannah} for extended values of $T_D$. Next to the largest difference between $P_T\left(\widetilde{\psi}^S_3\right)$ and $P_H\left(\widetilde{\psi}^S_3\right)$ at $T_D\approx5.0 \text{ \textmu s}=\pi/\delta$, similar splittings occur at $T_D\approx 4.6 \text{ \textmu s}$ and $T_D\approx 5.4 \text{ \textmu s}$. These durations are such that subsequent shift operations displace into opposite directions in the co-rotating phase space. That is, at odd step numbers of the QW the shift operation acts as $\ket{T}\ket{\alpha_k}\rightarrow\ket{T}\ket{\alpha_{k+1}}$ and at even steps as $\ket{T}\ket{\alpha_k}\rightarrow\ket{T}\ket{\alpha_{k-1}}$, and analogously for $\ket{H}\ket{\alpha_k}$. This creates again a QW in which the shift directions of $\ket{T}$ and $\ket{H}$ are exchanged at each step.
}
\label{scanweit}
\end{figure}

The duration $T_{D}$ of the dipole force pulses (and the related
mutual delays, see figure \ref{pulsschema}) is a very sensitive
parameter for the implementation of the QW. For a given detuning
$\delta$, $T_D$ determines the relative direction of subsequent
shifts in the co-rotating phase space. By altering $T_{D}$ we can
control and maximize the overlap of the interfering parts of the
wave function in the QW (Figure \ref{schrittidee}). We repeat the QW
pulse sequence with increasing values of $T_D$ (Figure
\ref{pulsschema}) and acquire the coin-state probabilities $P_{T}$
and $P_{H}$ via the coin-state detection (Figure \ref{scannah}). The
ratio $P_{T}/P_{H}$ is an indicator for the amount of interference.
If no interference occurs, it amounts to $P_{T}/P_{H}=1$. We
maximize the ratio by iterating to the optimal dipole force
duration, denoted as $T^{QW}_D\approx\pi /\delta$. The nonlinearity
of the dipole force, in particular the reduced return time discussed
in sect. \ref{dynamicsdescription}, leads to a deviation of
$T^{QW}_D$ from the duration $\pi/\delta$, optimal within the LDA
only. Additionally, this method is suitable to implicitly determine
the detuning $\delta$ to the required precision.

As illustrated in figure \ref{scannah}, the maximum ratio of the
measured probabilities amounts to $P_{T}/P_{H}\approx3$, with the
related dipole force duration being defined as $T^{QW}_D$. Averaged
over 60000 measurements the coin state probabilities at this point
amount to $P_{T} = 0.741 \pm 0.002$ and $P_{H} = 0.259 \pm 0.001$,
which is close to the theoretical predictions $0.75$ and $0.25$
(Figure \ref{schrittidee}). At slightly different values of the
dipole force duration, $T_D =T^{QW}_D\cdotp(1\pm 0.02)$ the coin
state probabilities are approximately equal ($P_{T}/P_{H}\approx1$),
indicating that the overlaps and hence the interference of different
parts of the wave function vanish.

The results of a numerical simulation of this procedure within 3SB
are in good agreement with the experimental data (Figure
\ref{scannah}). Additionally, the simulation shows similar
splittings of the coin state probabilities at the dipole force
durations $T_D=4.6 \mu s$ and $T_D=5.4 \mu s$, for
$\delta=2\pi\cdotp100\text{ kHz}$ (Figure \ref{scanweit}). These are
QWs in which the step directions of $\ket{T}$ and $\ket{H}$ are
exchanged at each step.

The simulation also shows a high frequent modulation of the coin
state probabilities in dependence of $T_D$. This is due to the fast
modulations of the trajectories in the rotating phase space, mainly
caused by $\ket{n+2}\leftrightarrow\ket{n}$-contributions of the
dipole force (See figure \ref{omegas}).
\begin{figure}
\centering
\includegraphics[width=12cm]{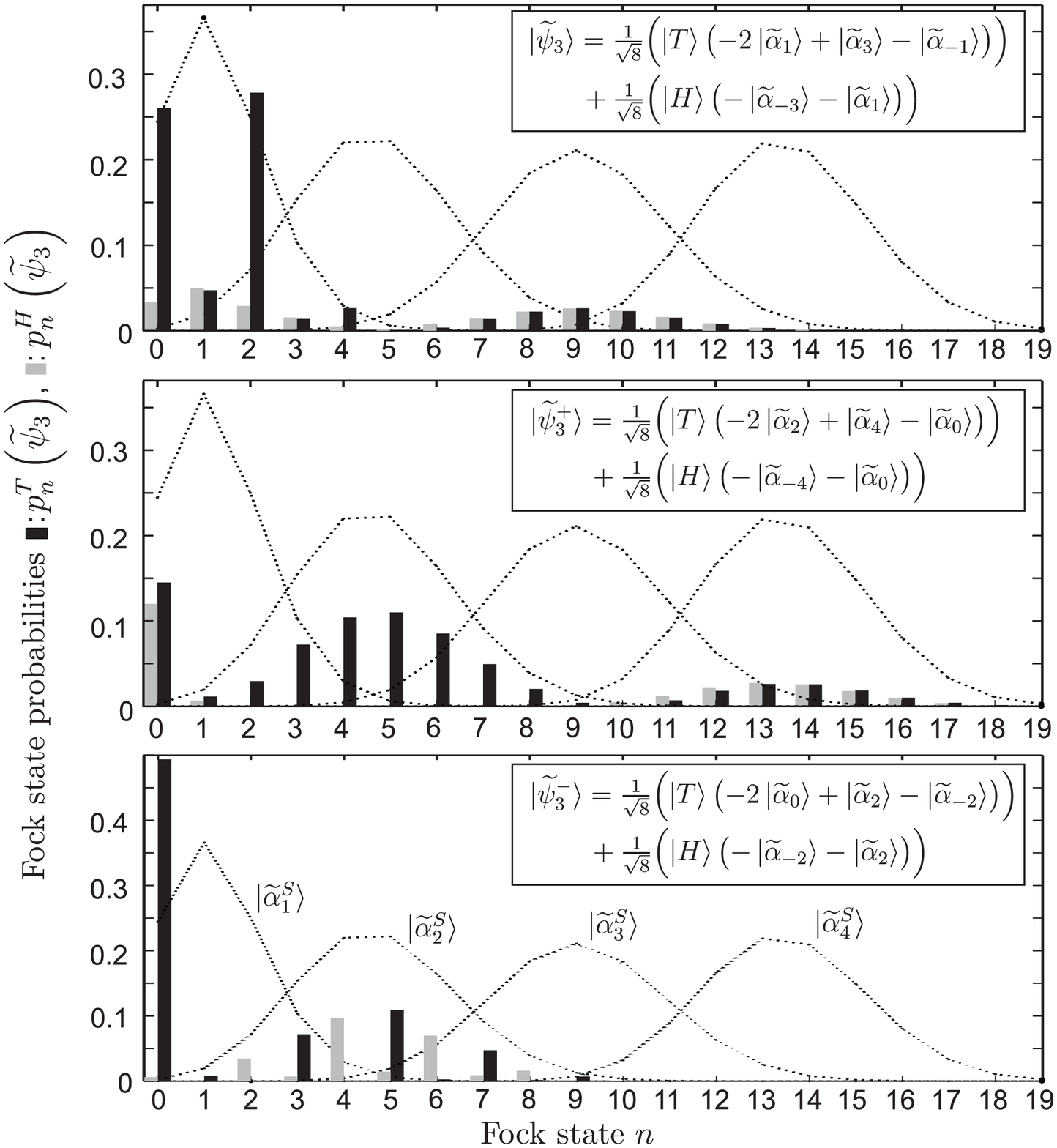}
\caption{Fock state probabilities $p^T_n(\psi)=\lvert\bra{T}\braket{n|\psi}\rvert^2$ and $p^H_n=\lvert\bra{H}\braket{n|\psi}\rvert^2$ of the simulated QW state $\ket{\widetilde{\psi}^{S}_3}$ (top), the shifted states $\ket{\widetilde{\psi}^{+,S}_3},\ket{\widetilde{\psi}^{-,S}_3}$ (center, bottom) and of the position states $\ket{\alpha^S_k}$ (dotted lines), as in figure \ref{positionen}.
All simulations are computed with equal parameters ($T^{QW}_D=4.990 \text{ \textmu s}$ and the 3SB approximation).
Each simulated state is in agreement with the corresponding experimental data, as exemplarily depicted in figure \ref{positionen} (inlay). Top: $\ket{\widetilde{\psi}_3}$ contains a superposition of $\posr{1}$ and $\posr{-1}$. This causes an interference (not related to the interference of the QW) in Fock space, resulting in high probabilities for even ($n=0,2,4$) and low probabilities for odd Fock states. With the states $\ket{\widetilde{\psi}^{+,S}_3}$ (center) and $\ket{\widetilde{\psi}^{-,S}_3}$ (bottom), where the position occupations are shifted by one position, it is possible to distinguish the probabilities corresponding to $\posr{k}$ and $\posr{-k}$.
}
\label{qw_verteilungen}
\end{figure}
\begin{figure}
\centering
\includegraphics[width=14cm]{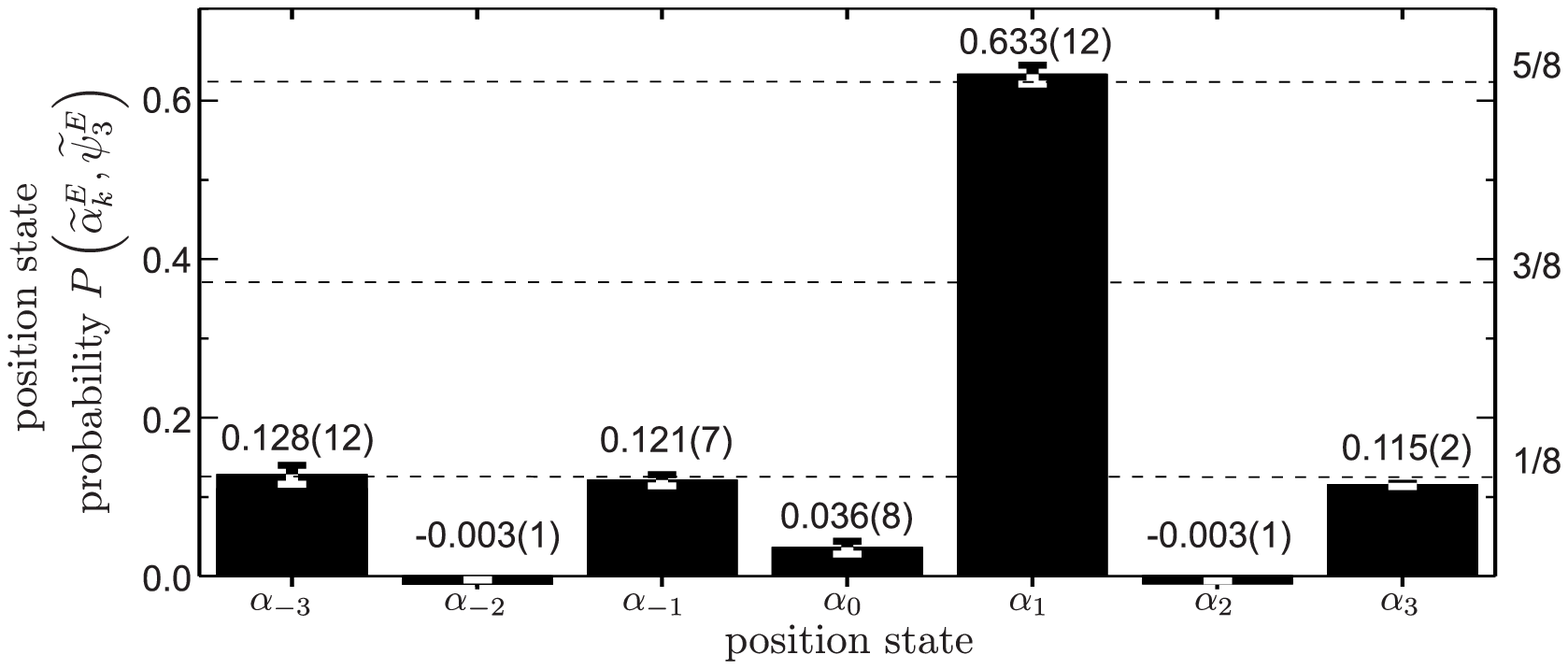}
\caption{Position probabilities $P(\widetilde{\alpha}^E_k,\widetilde{\psi}^E_3)$ after three steps of the asymmetric QW. These are acquired by a discrete Fourier analysis of the data from the motional state detection \eqref{bsbkurve} of the states $\widetilde{\psi}^E_3$, $\widetilde{\psi}^{+,E}_3$, $\widetilde{\psi}^{-,E}_3$ (Cf. figure \ref{qw_verteilungen}). The error bars represent the errors of the discrete Fourier analysis. The probabilities are in agreement with the theoretical values of an asymmetric QW (See figure \ref{schrittidee}). In particular the difference of the probabilities $P(\widetilde{\alpha}^E_1,\widetilde{\psi}^E_3)$ and $P(\widetilde{\alpha}^E_{-1},\widetilde{\psi}^E_3)$ indicates the high fidelity of the implementation of the QW.
The probabilities at the positions $\pose{-2}$, $\pose{0}$ and $\pose{2}$ remain nonzero due to the overlaps with the neighbouring position states. The probabilities due to the overlaps have been subtracted, using the probabilities at $\pose{\pm1}$ and $\pose{\pm3}$ as a reference. Therefore the remaining nonzero values are an indicator for the error of the implementation and readout of the QW.
}
\label{balkenbild}
\end{figure}

\section{Experimental results and conclusion}\label{conclusion}
We implement the QW pulse sequence (Figure \ref{pulsschema}) with
the optimized dipole force duration $T_D^{QW}$ (Figure
\ref{scannah}) and apply the motional state readout (Sect.
\ref{detection}). The resulting Fock state probabilities from the
corresponding simulation are illustrated in figure
\ref{qw_verteilungen}. To distinguish the position states $\posr{k}$
and $\posr{-k}$, which have the same Fock state probabilities, we
also apply the motional state readout to the states
$\ket{\widetilde{\psi}^+_3}$ and $\ket{\widetilde{\psi}^-_3}$, where
after the QW pulse sequence an additional shift operation towards
higher (lower) position states has been applied. The position state
probabilities corresponding to the experimental data are illustrated
in figure \ref{balkenbild}.

The experiment demonstrates the feasibility of implementing a QW
with a trapped ion. Although the number of steps is small in our
experiment, the trapped-ion system clearly reveals its strengths in
the high fidelity of the results. The severe limitation on the
number of steps for the implementation of the shift operator via the
optical dipole force is due to the Lamb--Dicke parameter $\eta$,
since shifts in terms of the displacement operator are only possible
within the LDR (Figures \ref{matrixelemente} and \ref{omegas}). Our
comparatively large Lamb--Dicke parameter $\eta=0.31$ allows for
three well distinguishable ($\lvert\Delta\alpha\rvert=1$) steps. For
a Lamb--Dicke parameter of $\eta=0.1$ the limit of the LDR is
$g_1=85$ with a corresponding maximal position state
$\ket{\alpha_{max}\approx9}$ within the LDR. This would allow for a
QW with $9$ steps of similar fidelity, using our scheme.
Equivalently, a Lamb--Dicke parameter of $\eta=0.06$ would allow for
15 steps. Additionally, since in such a setup the step size is very
small compared to the size of the LDR, the threshold $g_1$ may be
less of a limitation, as it can be overcome via small steps (Cf.
figure \ref{schritte}). A QW with up to 23 steps, implemented with a
dipole force on resonance ($\delta=0$), has recently been
demonstrated \cite{Zaehringer2010}. As described in section
\ref{theory}, a QW with effectively orthogonal position states
requires a step size of $\lvert\Delta\alpha\rvert\geq2$. This
reduces the number of steps within the LDR for any given setup.
Extending the number of possible steps substantially by further
reducing the Lamb--Dicke factor is a difficult task. The trap
frequency $\omega_z$ has to be increased and the mutual angles of
the laser beams providing the dipole force must be reduced (Cf.
figure \ref{dipolkraftschema}b). However, a small Lamb--Dicke factor
$\eta$ yields a weak coupling of the light field to the motional
degree of freedom. This weak coupling must be compensated by an
increased intensity of the laser beams. This in turn results in an
increased rate of spontaneous emission \cite{Wineland2003} from the
off-resonantly excited $\ket{^2P_{3/2},F=4}$ states (Figure
\ref{termschema}) and thus in a reduced coherence time for the QW.

\section{Outlook}\label{outlook}
\subsection{Implementation on the shift operator using photon kicks}
In the following we propose the implementation of the shift operator
with photon kicks \cite{Garc'ia-ripoll2003, Garc'ia-ripoll2005},
which is substantially less dependent on the motional state and
allows for the implementation of QWs with many steps. The principle
of a photon kick is to apply a $\pi$-pulse on the coin states which
is sufficiently short such that the free harmonic motion of the ion
during the pulse itself is negligible. It has been shown that the
change of the momentum of the ion during such a pulse can be
described by a displacement operator, allowing us to propose its
application as a building block for the shift operator of a QW. In
the original protocol \cite{Garc'ia-ripoll2003}, which has been
realized recently \cite{Campbell2010}, however, the influence of the
motional state on the performance of the photon kicks has not been
considered, since the amplitudes of the motional states were assumed
to remain small. For the implementation of a QW with many steps, we
have to consider (coherent) motional states with very large
amplitude and, thus, have to re-assess the validity of the
above-mentioned approximation. We find that, for a given fidelity,
the upper bound for the pulse duration scales inversely with the
motional amplitude, and additionally, for coherent motional states,
depends on the phase of their harmonic oscillation at the moment
when the pulse is applied. In the following we derive an analytic
bound for general states and present the results of a numerical
study for coherent motional states. With the latter we show that QWs
with up to 100 steps for a step size of $\lvert\Delta\alpha\rvert=2$
should be possible with state-of-the-art technology.

Referring to \cite{Garc'ia-ripoll2003}, we start our analysis with the Hamiltonian
\begin{equation}\label{starthamiltonian}
\begin{split}
\mathcal{H} &= \mathcal{H}_0+\mathcal{H}_1\\
&= \frac{\Omega}{2} \left( \sigma_+ \otimes e^{i\eta(a^\dagger+a)}+\sigma_- \otimes e^{-i\eta(a^\dagger+a)}\right)+\omega_z a^\dagger a.
\end{split}
\end{equation}

This Hamiltonian can be implemented in various ways, e.g. via direct
dipole coupling, two-photon stimulated Raman transitions or
stimulated Raman adiabatic passage \cite{Garc'ia-ripoll2003}. Each
implementation imposes different constraints on pulse duration,
laser intensities, etc. In the following we will focus on the
implementation with a two-photon stimulated Raman transition and
consider the energy levels of $^{25}\text{Mg}^+$.

In this configuration, two laser beams ($R_a$,$R_b$) resonantly
drive two-photon transitions between the coin states via a virtual
state detuned from the $P_{3/2}$ manifold by $\Delta_R$. Each laser
beam drives only one of the two Raman branches, due to their
different polarizations. In a RWA, terms varying at optical
frequencies are neglected. This is valid in our case for pulse
durations well above $1/10^{-15} \text{ Hz} = 1 \text{ fs}$.
Finally, an adiabatic elimination of the $P_{3/2}$ states requires
$\lvert \Omega/\Delta_R\rvert \ll 1$. The pulse duration $T_p$ in
our case must therefore be sufficiently longer than $5 \text{ ps}$
for $\Delta_R\approx 2\pi\cdotp10^{11} \text{ Hz}$ and
$T_p\Omega=\pi$ (see below). The effective wave vector of the
two-photon transition is $\bi k=\bi{k_a}-\bi{k_b}$.

Hamiltonian \eqref{starthamiltonian} implements the desired
displacement operator for a pulse duration of $T_p=\pi/\Omega$, if
we neglect the perturbation $\mathcal{H}_1$. The time evolution
operator then reads
\begin{equation}
\begin{split}
U_0(T_p)=&e^{-i\mathcal{H}_0T_p}\\
=&\cos\left(\frac{\Omega T_p}{2}\right)\cdotp\idty_{coin}\otimes\idty_{motion}\\
&-i\sin\left(\frac{\Omega T_p}{2}\right)\cdotp\left(\sigma_+\otimes D\left(i\eta\right)+\sigma_-\otimes D\left(-i\eta\right)\right)\\
=&-i\left(\sigma_+\otimes D\left(i\eta\right)+\sigma_-\otimes D\left(-i\eta\right)\right),
\end{split}
\end{equation}
which is obtained by expanding the exponential function, splitting
the series into odd and even parts and using the properties of the
Pauli matrices and displacement operators.

The shift operator itself, implementing the desired step size (i.e.
$\lvert\Delta\alpha\rvert=2$, see figure \ref{qw_schmelz}), can be
realized by the subsequent application of $2/\eta$ kicks in such a
way that the displacements $D(i\eta)$ of several $\pi$-pulses add up
to $D(\Delta \alpha=i\eta\cdotp2/\eta)$. This can be achieved by
changing the direction of the effective wave vector by $180^{\circ}$
for each photon kick. In practice one can either switch between two
Raman beam configurations with opposite effective wave vectors, or
implement every second $\pi$-pulse by a RF transition for which the
momentum transfer is negligible. Notably, with this protocol the
step sizes for both directions of the QW are equal, in contrast to
the method of optical dipole forces used in our current experiment.

In the following we derive a conservative estimate for the deviation
from a coherent-state displacement induced by $\mathcal{H}_1$. The
total time evolution is
\begin{equation}
U(t)=e^{-i\mathcal{H}t}= U_0(t)\cdotp V(t)
\end{equation}
with $V(t)= e^{i\mathcal{H}_0t}e^{-i\mathcal{H}t}$. $V(t)$ can be differentiated
\begin{equation}
\dot{V}(t)=-ie^{i\mathcal{H}_0t}\mathcal{H}_1e^{-i\mathcal{H}_0t}\cdotp V(t),
\end{equation}
leading to an equation wich is formally solved by the integral equation
\begin{equation}
V(t)=\idty-i\int^t_0\rmd s \;\, e^{i\mathcal{H}_0s}\mathcal{H}_1e^{-i\mathcal{H}_0s}\cdotp V(s),
\end{equation}
using $V(0)=\idty$. Now we define $\epsilon$ as the distance between
the evolved state according to the full Hamiltonian and the desired
evolved state according to $\mathcal{H}_0$:
\begin{equation}
\begin{split}
\epsilon&\equiv\|\left(U(t)-U_0(t)\right)\ket{\psi}\|\\
&=\|\left(V(t)-\idty\right)\ket{\psi}\|\\
&=\| \int^t_0\rmd s \;\, e^{i\mathcal{H}_0s}\mathcal{H}_1e^{-i\mathcal{H}_0s} V(s) \ket{\psi}\|.
\end{split}
\end{equation}
Approximating the last expression by the largest term of the first
order Dyson series, and considering a motional state
$\ket{\psi}=\ket{H}\ket{\alpha}$, gives the following error estimate
for a pulse with duration $T_p$ \cite{Garc'ia-Ripoll3}
\begin{equation}
\epsilon\approx\|\int_0^{T_p}\rmd s \;\, \omega_z\idty\otimes a^\dagger a \ket{\psi}\|= T_p\omega_z\lvert\alpha\rvert^2.
\end{equation}
Thus, for an initial state $\ket{H}\ket{\alpha}$ (the coin state can
be chosen arbitrarily), the pulse duration $T_p$ necessary to
implement the displacement operator with an error smaller than
$\epsilon$ must fulfill
\begin{equation}
T_p\leq\frac{\epsilon}{\omega_z\cdotp\lvert\alpha\rvert^2}.
\end{equation}
The scaling with $\lvert\alpha\rvert^{-2}$ is, however, a rather
rough estimate. This is shown by a numerical simulation of this
process, in particular considering the application of photon kicks
to (superpositions of) coherent motional states. We compute the
fidelity
$f=\lvert\bra{H}\bra{\alpha}U_0^\dagger(T_p)U(T_p)\ket{H}\ket{\alpha}\rvert^2$
with the initial state $\ket{H}\ket{\alpha}$, a pulse duration
$T_p$, and $\Omega=\pi/T_p$, where the time evolution is implemented
using a Runge--Kutta method. The results show that the fidelity
strongly depends on the phase of ion oscillation at the moment of
the photon kick.

Demanding a fidelity of $f\ge 0.99$ and for imaginary $\alpha$, i.e.
at the moment of the photon kick the ion is in the center of the
harmonic potential and thus fastest, for our experimental parameters
we find \footnote{Equations \eqref{imagnumerics} and
\eqref{realnumerics} are the results of quadratic fits to
double-logarithmic plots of pairs ($T_p$, $\lvert\alpha\rvert$) for
a fidelity $f=0.99$ and $\lvert\alpha\rvert\leq10$. Higher motional
amplitudes were not considered due to sizable additional numerical
effort. For the following estimates, the scaling is considered to be
preserved.}
\begin{equation}\label{imagnumerics}
T^{\Im}_{f=0.99}(\lvert\alpha\rvert)=\exp\left(-17.55-0.63\ln(\lvert\alpha\rvert)-0.05\left(\ln\left(\lvert\alpha\rvert\right)\right)^2\right).
\end{equation}
For $\lvert\alpha\rvert=200$, an amplitude reached after the 100th
step of a QW with $\lvert\Delta\alpha\rvert=2$, the pulse duration
must be shorter than $T^{\Im}_{f=0.99}(200)=0.21\text{ ns}$.

However, applying the photon kick when the ion is at its turning
point, i.e. the ion is slowest and $\alpha$ is real, the scaling is
less demanding. We find
\begin{equation}\label{realnumerics}
T^{\Re}_{f=0.99}(\lvert\alpha\rvert)=\exp\left(-17.03-0.02\ln(\lvert\alpha\rvert)-0.1\left(\ln\left(\lvert\alpha\rvert\right)\right)^2\right).
\end{equation}
Most importantly, the prefactor of the term linear in
$\ln(\lvert\alpha\rvert)$ is much smaller than for an imaginary
$\alpha$. For the 100th step, the pulse duration therefore only has
to be shorter than $T^{\Re}_{f=0.99}(200)=2.18\text{ ns}$, which is
within the specifications of a fast-switching electro-optic
modulator and our current continuous-wave laser system. Timing the
application of the photon kick to the (spatial) turning points of
all the coherent oscillations occuring during the QW is possible,
because we start the QW in the motional ground state and the
position states are aligned along a line in the co-rotating phase
space. Thus, coherent states of different $\lvert\Delta\alpha\rvert$
reach their turning points simultaneously.

Given our experimental parameters, in particular the width of the
ground state wave function $z_0=10\text{ nm}$, the coherent motional
state of maximal amplitude, $\ket{\alpha_{max}=200}$, would have a
real-space amplitude of
$\bra{\alpha_{max}}z\ket{\alpha_{max}}=4\text{ \textmu m}$. At such
high motional amplitudes anharmonicities of the trapping potential
must be considered. These depend on the design of the electrodes and
could be eliminated, e.g. by designing the Paul trap electrodes in a
hyperbolic shape \cite{Paul1990}. Additionally, micromotion might
increase the deviation from the ideal walk, for example by reducing
the overlap of the additionally oscillating motional wave functions.
However, it will remain negligible when the QW is implemented in the
axial degree of freedom of an ion in a linear Paul trap.

\subsection{QW in higher dimensions}
A QW in two or three dimensions is possible by additionally
considering the motion in the radial direction. The pulse sequence
for a step of a QW is then the subsequent application of the shift
operator in each direction where each operation is preceeded by a
coin toss.

More possibilities and reduced technical requirements might be
achieved by trapping more than one ion and considering the
collective degrees of motion in one direction. Reference
\cite{Omar2006} describes the scheme with two ions, creating a
4-sided coin, where two coin states affect the walk in the
center-of-mass motional mode and the other two in the stretch mode
of motion. In particular, possibilities with the coin being
initialized in an entangled state are investigated.

A photon kick, as decribed above, induces motion in all motional
modes in the direction of the effective wave vector $\bi{k}$,
according to the respective coin states. That is, with $N$ ions, one
step of the QW consists of a coin operation on the $2^N$-sided coin
and a single shift operation, that displaces the part of the
motional wave function related to each coin state into opposite
direction in phase space of one motional mode. The particular
difficulty is to assign the $2^{N-1}$ pairs of coin states to $N$
different axial motional modes such that for each coin state the
corresponding state dependent force induces motion in one direction
in one certain mode \cite{James1998}. One possibility to obtain the
required number of motional modes is to add ions that do not contain
a transition corresponding to the coin states and therefore are not
affected by the photon kicks. That would be in our case
$^{24}\text{Mg}$ ions without hyperfine structure and therefore no
coin states and no corrsponding transition. With this, the
implementation of a QW in four dimensions is possible using three
$^{25}\text{Mg}$ ions and one $^{24}\text{Mg}$ ion. For more
dimensions the issue arises that the $^{24}\text{Mg}$ and
$^{25}\text{Mg}$ ions have to be arranged in such a way that for
each coin state the corresponding state dependent force induces
motion in one certain mode, which has not been clarified yet.

\section*{Acknowledgements}
This work was supported by MPQ, MPG, DFG (SCHA 973/1-6), the EU via
SCALA and STReP PICC, and the DFG Cluster of Excellence
``Munich-Centre for Advanced Photonics''. RM acknowledges support by
the EU project COQUIT. We thank J. J. Garc\'{i}a-Ripoll for
intriguing discussions about the short-pulse scheme, M. J. McDonnell
for assistance with the simulation code, and Ignacio Cirac, Gerhard
Rempe and Reinhard Werner for their great intellectual and financial
support.

\section*{References}

\bibliography{bibliography}
\bibliographystyle{iopart-num}
\end{document}